\documentclass[twocolumn,showpacs,superscriptaddress,preprintnumbers,prd]{revtex4}
\usepackage{graphicx,amsmath,dcolumn,bm}
\usepackage{mathrsfs}
\usepackage{graphicx}
\usepackage{epstopdf}
\usepackage{epsfig}
\usepackage{pdfsync}
\newcommand{\ra}{\rightarrow}
\def\lsim{\mathrel{\rlap{\lower4pt\hbox{\hskip1pt$\sim$}}
    \raise1pt\hbox{$<$}}}
\def\slash#1{\hspace*{-0.05cm}\not\!#1}

\newcommand{\slashes}[1]{#1\hskip-0.5em /}
\usepackage{color}

\definecolor{mycolor}{rgb}{0.6,0.0,0.4}

\begin{document}
\preprint{KEK-TH-1224}
\title{\Large \bf 
Novel two-to-three hard hadronic processes and \\
possible studies of generalized parton distributions at hadron facilities}
\author{S. Kumano}
\affiliation{Institute of Particle and Nuclear Studies \\
          High Energy Accelerator Research Organization (KEK) \\
          1-1, Ooho, Tsukuba, Ibaraki, 305-0801, Japan}
\affiliation{Department of Particle and Nuclear Studies \\
             Graduate University for Advanced Studies \\
           1-1, Ooho, Tsukuba, Ibaraki, 305-0801, Japan}     
\author{M. Strikman}
\affiliation{Department of Physics, Pennsylvania State University \\
             University Park, PA 16802, U.S.A.} 
\author{K. Sudoh}
\affiliation{Institute of Particle and Nuclear Studies \\
         High Energy Accelerator Research Organization (KEK) \\
         1-1, Ooho, Tsukuba, Ibaraki, 305-0801, Japan}
\affiliation{Nishogakusha University,
         6-16, Sanbancho, Chiyoda, Tokyo, 102-8336, Japan}
\date{September 4, 2009}

\begin{abstract}
We consider a novel class of hard branching hadronic processes
$a+b \to c+d +e$, where hadrons $c$ and $d$ have large and nearly opposite
transverse momenta and large invariant energy which is a finite fraction
of the total invariant energy.  We use color transparency logic to 
argue that these processes can be used to study quark generalized
parton distributions (GPDs) for baryons and mesons in hadron collisions 
hence complementing and adding to the studies of GPDs in the exclusive DIS 
(deep inelastic scattering) processes. We propose that a number of GPDs
can be investigated in hadron facilities
such as J-PARC (Japan Proton Accelerator Research Complex) facility
and GSI-FAIR (Gesellschaft f\"ur Schwerionenforschung -Facility for 
Antiproton and Ion Research) project. In this work, the GPDs for the nucleon
and for the $N \rightarrow \Delta$ transition are studied in the reaction 
$N+N \rightarrow N+\pi+B$, where $N$, $\pi$, and $B$ are a nucleon,
a pion, and a baryon (nucleon or $\Delta$), respectively, with a large
momentum transfer between $B$ (or $\pi$) and the incident nucleon.
In particular, the Efremov-Radyushkin-Brodsky-Lepage (ERBL) region
of the GPDs can be measured in  such exclusive reactions.
We estimate the cross section of the processes $N+N \rightarrow N+\pi+B$
by using current models for relevant GPDs and information about 
large angle $\pi N$ reactions. We find that it will be feasible 
to measure these cross sections at the high-energy hadron facilities,
and get novel information about the nucleon structure, for example,
contributions of quark orbital angular momenta to the nucleon spin.
The studies of $N \rightarrow \Delta$ transition GPDs
could be valuable also for investigating electromagnetic
properties of the transition.
\end{abstract}

\pacs{13.85.-t, 13.60.Le, 12.38.-t}
\maketitle

\section{Introduction}
Understanding complexity of the structure of hadrons is one 
of prime objectives of QCD (Quantum Chromodynamics). For a long time,
the main focus was the study of single parton densities 
in inclusive processes for which QCD factorization theorem is
valid \cite{Collins:1989gx}. More recently, many investigations
started to focus on nucleon spin structure
by polarized lepton-nucleon scattering and proton-proton collisions.
It became clear that only a small fraction (20$-$30 \%) of nucleon spin
is carried by quarks \cite{ppdfs}. There are still a large uncertainty
in the gluon spin contribution \cite{ppdfs}; however, recent measurements
indicate a tendency that it is also small in the region of momentum
fraction $x \sim 0.1$ \cite{deltaG}. 
It is, therefore, important to find a contribution from
orbital angular momenta as a possible source for explaining
the nucleon spin.

A parallel development of the last two decades was realization that
the QCD factorization theorem is valid for deeply virtual Compton
scattering \cite{DVCSfact} and for exclusive production of mesons 
in deep inelastic scattering (DIS) by longitudinally polarized photons
\cite{Brodsky:1994kf,Collins:1997hv}. Amplitudes of these processes are
expressed through universal objects - generalized parton distributions
(GPDs) which contain information about both longitudinal and transverse
distributions of hadrons. 
 
In the forward scattering limit, several GPDs become usual
parton distribution functions with the transverse momentum dependence
connected to the transverse coordinate distribution of partons
in the hadron \cite{transverse}. Therefore, studies of GPDs allow to 
establish the three-dimensional image of the nucleon and hence provide
critical information for description of nucleon-nucleon collisions
at collider energies \cite{annual}. The first moments of GPDs
are given by the corresponding hadron form factors 
providing a link between high-energy and medium-energy dynamics. 
At the same time the second moment of a certain combination
of GPDs is related to the total angular momentum contribution
to the nucleon spin and hence probes orbital angular momentum effects
in the spin structure of nucleon \cite{ji97}.

For the recent reviews of the properties of GPDs and applications
to the study of the hard exclusive DIS processes, we refer the reader
to Refs. \cite{gpv01,diehl03}. 
Several theoretical calculations have been performed 
\cite{bochum,osaka,chiral-p-08,lattice,aligned-jet,gpd-paramet}
and we will use the ones described in Ref. \cite{bochum}.
Experimental studies of GPDs are
performed at HERA (Hadron-Elektron-Ringanlage) collider and 
in several fixed target experiments \cite{gpd-exp,recent-gpds}.
It was also proposed to study GPDs in
the $\pi N \rightarrow \mu^+ \mu^- +N$ process \cite{gpd-pp-l+l-}.

The QCD factorization theorem for exclusive hard processes induced
by longitudinally polarized photons is valid
for any exclusive final states in particular exclusive
meson production \cite{Collins:1997hv}. Hence, 
one can probe various non-diagonal transitions between baryons 
- transitional GPDs \cite{fpsv9800}.  
For example, the GPDs for the $N \rightarrow \Delta$
transition, where $N$ and $\Delta$ indicate the nucleon 
and $\Delta (1232)$, should be important for the studies of
nucleon spin \cite{fpsv9800}.
There is also an important link to the $N\to \Delta$ transitional
electromagnetic form factors, which are studies
in the low-energy electron experiments \cite{NDelta-form}. 
Electromagnetic moments of unstable particles
such as the $\Delta$ cannot be measured by usual methods
for stable nuclei \cite{hkmm87}.
Electromagnetic moments for the $N \rightarrow \Delta$ are 
interesting quantities \cite{NDelta-form}. In particular,
the C2 and E2 transition moments indicate shape deformations of
the nucleon and $\Delta$, so that they are valuable for proving
interesting information on quark-gluon dynamics \cite{NDelta-E2}.
The $N \rightarrow \Delta$ GPDs are related to the nucleonic GPDs
according to a chiral-quark soliton model in the large $N_c$ limit
\cite{fpsv9800,gpv01}, where $N_c$ is the number of colors.
Using the relation to the nucleonic GPDs, we could learn about
contributions of quark orbital angular momenta to the nucleon spin.

It was suggested in Refs. \cite{stardust,Frankfurt:2002kz} that one can
investigate presence of small-size color singlet $q\bar q$ and 
$qqq$ clusters in hadrons using large-angle branching hadronic
processes $a+ b\to c + d + e$ where the hadron $e$ is produced
in the fragmentation of $b$ with fixed Feynman $x_F$ 
and fixed transverse momentum $p_T^{(e)}$, while 
the hadrons $c$ and $d$ are produced with large and near balancing
transverse momenta: $p_T^{(c)} \approx - p_T^{(d)}$.

In this paper, we will argue that investigations of these processes
allow to probe various diagonal and transitional GPDs of hadrons.
We will focus on one of the simplest  subclass of these reactions 
$N+N \rightarrow N+\pi+B$, where $\pi$ indicates the pion and $B$
does the nucleon or $\Delta$. We will argue that these reactions 
allow to probe the GPDs for the nucleon and the $N \rightarrow \Delta$
transition in the kinematical region of 
Efremov-Radyushkin-Brodsky-Lepage (ERBL) \cite{erbl}
at hadron facilities in addition to the lepton scattering measurements.
Especially, there are future projects such as
the J-PARC (Japan Proton Accelerator Research Complex) facility
\cite{j-parc} and GSI-FAIR (Gesellschaft f\"ur Schwerionenforschung 
-Facility for Antiproton and Ion Research) project \cite{gsi-fair}
for investigating high-energy proton and anti-proton reactions.
Other branching reactions which one can study at these facilities
include reactions initiated by meson projectiles which provide
a unique way to probe meson GPDs: 
$\pi +N \to \pi + \pi + \mbox{low-}p_T \ \mbox{baryon}$
and processes initiated by protons (antiprotons)  involving production 
of two large angle baryons (baryon and antibaryon at the GSI-FAIR)
with a small transverse momentum for the pion:
$p(\bar{p}) +N \to p(\bar{p}) + N + \mbox{low-}p_T \ \mbox{meson}$.
The cross sections of the processes with two large transverse
momentum baryons are expressed by skewed (transitional) parton
distributions $N\to M$ \cite{Frankfurt:2002kz}, where $M$
indicates a meson. 
Experimental indications for existence of back to back production
of two protons with large $p_t$ were reported in 
Ref. \cite{Zhalov:2000hk}.

This article is organized as follows.
In Sec. \ref{cross-section}, we provide arguments for validity 
of factorization for $two \to three $ process and introduce
the description of the $N+N \rightarrow N+\pi+B$ cross section
in terms of a $N \rightarrow B$ amplitude
and a meson-nucleon scattering part. 
The GPDs for the nucleon and the $N \rightarrow \Delta$ transition
are defined and their major properties are introduced in Sec. \ref{gpds}.
The $N \rightarrow B$ amplitude is expressed through the nucleonic
or $N \rightarrow \Delta$ GPDs in Sec. \ref{n-b-amplitude}.
The meson-nucleon scattering cross sections are parametrized
so as to explain experimental measurements in Sec. \ref{piN}.
The $N+N \rightarrow N+\pi+B$ cross sections are shown
in Sec. \ref{results} by assuming a J-PARC kinematics 
for the initial proton and by taking a typical model
for the GPDs. The results are summarized in Sec. \ref{summary}.

\section{Factorization of cross section 
         for $\mathbf{N+N \rightarrow N + \pi + B}$}
\label{cross-section}

It was argued in Ref. \cite{bf7375} that cross sections of large angle
scattering processes: $a +b \to c+d$ in the limit of 
$s\to \infty$ with $\theta_{cm}$=const,
where $s$ is the center-of-mass (c.m.) energy squared and
$\theta_{cm}$ is the scattering angle,              
are dominated by the  short-distance configurations in hadrons both
in the initial and final states and satisfy dimensional counting rules. 
The data on a variety of such processes
(see Ref. \cite{White94} and references therein) 
are consistent with the dimensional counting rules. 
They also indicate that the processes, where quark exchange is allowed,
are much larger than those where it is forbidden. 
The especially interesting is the observed  systematics of reactions
induced by mesons: the ratio
of $K^+p\to K^+p$ and $\pi^+p\to \pi^+p$ cross sections
at $\theta_{cm}=90^o$, where $p$ is the proton,
is close to the square of the wave functions
$K^+$ and $\pi^+$ mesons at the origin:
$\left[f_K/f_{\pi}\right]^2\approx 1.4$, 
while it is $\sim 0.5$ for a small momentum transfer $|t|$.

A test of dominance of the small-size configurations 
in these processes was suggested by Brodsky and Mueller 
\cite{Brodsky,Mueller}. They pointed out that since small-size
configurations weakly interact with other hadrons the cross section
of the large angle processes like $a + A \to a + p +(A-1)$,
where $a$ and $A$ indicate a hadron and a nucleus with mass number $A$,
should be proportional to the number of protons in the nucleus.
This regime is usually referred to as the color transparency (CT).
The phenomenon which slows down the onset of the CT regime is
the space-time evolution of the small wave packages.
Evolution occurs over the distance scale $l_{coh}$ (coherence length), 
which can be estimated as 
$l_{coh}\sim 0.3 \div 0.4 \ {\rm fm} \cdot  p_h / {\rm GeV}$,
where $p_h$ is the hadron momentum \cite{Frankfurt:1994hf}. 
Recently, the color transparency phenomenon was observed in
the electroproduction of pions \cite{:2007gqa} with the change
of the transparency consistent with the prediction of
Ref. \cite{Larson:2006ge}, in which above estimate of 
$l_{coh}$ was used. For the review of the CT phenomena
in the high-energy processes, see Ref. \cite{annual}.
 
Overall, it appears that there is a good evidence for dominance
of the small-size configurations in the large angle meson-nucleon
scattering and for suppression of the interaction of small-size 
configurations near the interaction point. 
Hence, the meson-nucleon scattering at large angles should be
a good probe of short-distance effects already at moderate energies
$s_{\pi N} \sim 12$ ${\rm GeV^2}$ and  large $\theta_{cm} \sim 90^o$.
Therefore, we start our investigation of the $two\to three $ processes
with the reaction $N+N \rightarrow N + \pi + B$ in the kinematics
where nucleon scatters at large $|t'|$ off a small-size color
singlet $q\bar q$ clusters in the nucleon.
This is analogous to the process of $\gamma^* + N\to \pi + B$ 
scattering at $x\ge 0.2$, where dominant contribution originates
from the scattering off small-size color singlet $q\bar q$ clusters
in the nucleon \cite{gpv01}. These processes allow 
to investigate the nucleonic and 
$N \rightarrow \Delta$ transition GPDs. An additional advantage
is that these GPDs are extensively studied in the literature so
that we will be able to use existing analyses of these GPDs to 
perform numerical estimates. In this section, we explain how
to calculate the cross section of these reactions and give
arguments for validity of the factorization. 

\begin{figure}[b]
\begin{center}
\includegraphics[width=0.40\textwidth]{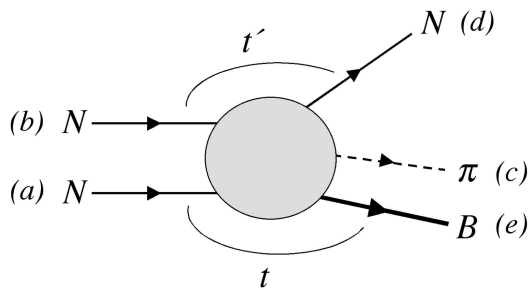}
\vspace{-0.3cm}
\caption{$N+N \rightarrow N + \pi + B$ reaction.}
\label{fig:diagram1}
\end{center}
\end{figure}

The cross section is given by
\begin{align}
d\sigma = & \frac{S}{4 \sqrt{(p_a \cdot p_b)^2-m_N^4}} 
          \overline{\sum_{\lambda_a, \lambda_b}} 
          \sum_{\lambda_d, \lambda_e}
          \left| {\cal M}_{NN\ra N\pi B} \right| ^2 
\nonumber \\
&  \times \frac{1}{2E_c}  \frac{d^3p_c}{(2\pi)^3}
          \frac{1}{2E_d} \frac{d^3p_d}{(2\pi)^3}
          \frac{1}{2E_e} \frac{d^3p_e}{(2\pi)^3}
\nonumber \\ 
&  \times  (2\pi)^4 \delta^4 (p_a+p_b-p_c-p_d-p_e) ,
\label{eqn:Xsection}
\end{align}
where ${\cal M}_{NN\ra N\pi B}$ is the matrix element
for the process $NN\ra N\pi B$, $m_N$ is the nucleon mass,
and $\lambda_i$ ($i$=$a$, $b$, $d$, $e$) is a spin of an initial
or final particle. 
The factor $S$ is $1/2$ if the final baryon is identical to
the final nucleon $N$ and it is 1 if they are not identical.
The particle indices $a$, $b$, $c$, $d$, and $e$
are assigned to the initial nucleons and final nucleon, pion, and 
baryon $B$ (nucleon or $\Delta$) as shown in Fig. \ref{fig:diagram1}.
Their momenta and energies are denoted as $p_i$ and $E_i$
($i$=$a$, $b$, $c$, $d$, $e$).
Averages are taken for the initial spins $\lambda_a$ and $\lambda_b$.
Mandelstam variables $s$, $t$, $s'$, and $t'$ are defined
\begin{alignat}{2}
s & =(p_{a}+p_{b})^2 ,  & \ \ \   t= & (p_{a}-p_{e})^2 , 
\nonumber \\
s' & = (p_c+p_d)^2,     & \ \ \   t'= & (p_{b}-p_{d})^2 ,
\label{eqn:mandelstam1}
\end{alignat}
for this reaction.
Note that two complementary kinematic regions contribute to
the discussed cross section. In one kinematics $B$ has small
momentum relative to nucleon (a) - the target nucleon, 
while in the second kinematics $t$ is small relative 
to the projectile - nucleon (b). These two cross sections
are obviously equal. The corresponding amplitudes do not
interfere in the discussed limit. Hence for simplicity 
we will present the results for one kinematics only. 
At the moment, there exists no procedure for calculating cross sections 
of $N+N \rightarrow N + \pi + B$ process in a generic kinematics.
However, it is simplified for the hard regime: $|t'| \gg m_N^2$
and $|u'|=|(p_b-p_c)^2| \gg m_N^2$.  
Formal arguments can be developed for the limit  
\begin{equation}
s' \to \infty, 
\ \ \frac{t'}{s'} ={\rm const},
\label{asympt}
\end{equation}
since the leading-order QCD diagrams dominate the cross section
of two-body processes \cite{bf7375}.
All the partons which experience a large change
of the direction have to be attached to hard lines.
This idea is used as a counting rule for estimating energy dependence
of exclusive cross sections at high energies \cite{bf7375,chh97}. 
In our studies, the variable $|t|$ is fixed and sufficiently 
small ($|t| \ll m_N^2$), and the center-of-mass energy is 
$\sqrt{s} \simeq \sqrt{60} \div  \sqrt{100}$ GeV at the J-PARC.

\begin{figure}[b]
\begin{center}
\includegraphics[width=0.40\textwidth]{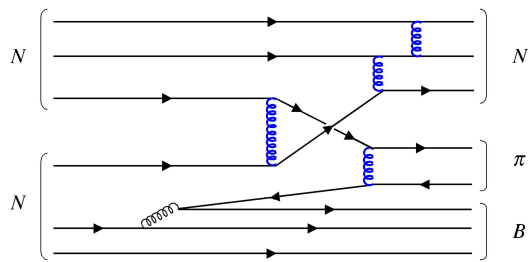}
\caption{(Color online) A typical leading subprocess contribution,
         which corresponds to Fig. \ref{fig:diagram2}, to
         the $N+N \rightarrow N+\pi+B$ cross section.
         The spirals indicate gluon exchanges. The thick spirals
         show hard gluon exchanges, whereas the narrow one shows 
         a soft gluon exchange.
         In the large $|t'|$ and $|u'|$ process,
         the quarks in the final baryon and pion should be attached
         to the hard gluon lines.}
\label{fig:subproc-a}
\end{center}
\end{figure}

Similar to the logic of the QCD counting rules of Ref. \cite{bf7375},
we separate diagrams with minimal number of hard gluons.
The hard part of these diagrams has the same topologic structure
as the diagrams for the corresponding elementary reaction 
$M + N \to M' +B$. A typical diagram of this kind is presented
in Fig. \ref{fig:subproc-a}.

\begin{figure}[t]
\begin{center}
\includegraphics[width=0.40\textwidth]{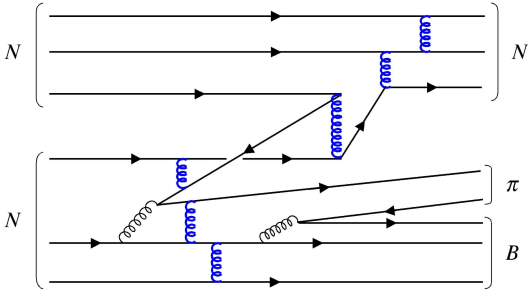}
\caption{(Color online) A typical subleading contribution
          to the $N+N \rightarrow N+\pi+\Delta$ cross section.
          As explained in the text, the nucleon $(a)$ splits as
          $N \rightarrow q\bar q  +B^* 
          \rightarrow \pi+B$, then
          the $q\bar q$ pair interacts with another nucleon $(b)$
          ($N+q\bar q \rightarrow N$). Notations for soft and hard
          gluons are the same in Fig. \ref{fig:subproc-a}.}
\label{fig:subproc-b}
\end{center}
  \vspace{-0.4cm}
\end{figure}

It is easy to check that other processes which contribute
to the $N+N \rightarrow N + \pi + B$ cross section, 
are suppressed by powers of $s'$ in the limit of Eq. (\ref{asympt}). 
For example, in the case of a diagram shown in 
Fig. \ref{fig:subproc-b}, the process proceeds via 
$N \rightarrow q\bar q+B^* \rightarrow q\bar q+\pi+B$,
where $B^*$ indicates an intermediate baryonic state,
in the lower blob of this figure and subsequent transition 
$N+q\bar q \rightarrow N$ in the upper one. In this case,
the amplitude is small because the $q\bar q$ propagation is 
suppressed by the factor $1/|t'|$ ($|t'| \gg m_N^2$), whereas
the  corresponding factor  in Fig. \ref{fig:subproc-a}
is $1/|t|$ ($|t| \ll m_N^2$). Furthermore, the large momentum ($|t'|$)
is also involved in the lower blob ($N \rightarrow q\bar q+B^*$)
in addition to the upper one.
It means that more hard gluon exchanges are involved in this case,
which suppresses the contribution in Fig. \ref{fig:subproc-b}.

Next, we need to demonstrate that additional soft interactions,
which are not suppressed by powers of $t'$, are canceled out.
The easiest way is to consider the process in the rest frame 
of fragmenting nucleon. Neglecting space-time evolution of the wave 
packets, which is legitimate in the limit of Eq. (\ref{asympt}), 
we observe that transverse size of the projectile and
two outgoing hadrons near the interaction point is 
given by $\sim 1/\sqrt{|t'|}$. Hence, the color
transparency arguments essential for the proof of factorization 
for exclusive hadron production in DIS \cite{Collins:1997hv} are
applicable, and the soft interactions are suppressed by at least
a power of $1/t'$.  
Hence, we find that the leading contribution to 
the $N+N \rightarrow N + \pi + B$ amplitude  is
given by the factorized form. 

\begin{figure}[t]
\begin{center}
\includegraphics[width=0.40\textwidth]{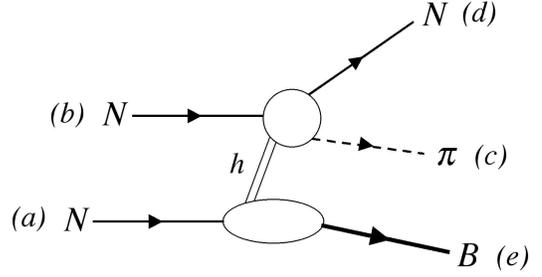}
\caption{Factorization of $N+N \rightarrow N + \pi + B$
         process into $N \rightarrow h+B$ and 
         $h+N \rightarrow \pi N$ amplitudes.}
\label{fig:diagram2}
\end{center}
  \vspace{-0.4cm}
\end{figure}

The transition amplitude ${\cal M}_{NN\ra N\pi B}$ is written as
a convolution of generalized parton distribution $G(N\to B)$ 
of $q \bar q$ pairs in the nucleon with a hard interaction blob $H$
of $q\bar q$ - $3 q$ elastic scattering and 
$q\bar q \ (3q)$ wave functions of the initial
nucleon $(b)$, the final nucleon $(d)$, and $(c)$:
\begin{equation}
{\cal M}_{N N \rightarrow N\pi B} 
= G(N\to B) \otimes \psi_N^{i} \otimes H \psi_{\pi}\otimes \psi_N^{f}.
\label{eqn:factorization1}
\end{equation}
Here, $\otimes$ denotes convolution of the different blocks, 
namely an integration over the light-cone fractions of
constituents involved in the process, $etc.$ 
There is also an implicit summation over the intermediate states.
We can also rewrite Eq. (\ref{eqn:factorization1}) in somewhat more 
transparent form as:
\begin{equation}
{\cal M}_{N N \rightarrow N\pi B} 
= {\mathscr M}_{N \rightarrow h B} 
  \, {\mathscr M}_{h N\ra \pi N} ,
\label{eqn:factorization}
\end{equation}
where $h$ indicates the intermediate $q\bar q$ hadronic state
and the nucleon, respectively. This corresponds to the diagram
of Fig.~\ref{fig:diagram2}.

Consequently, the cross section has the following generic
factorized form in the limit of Eq. (\ref{asympt}): 
 \begin{equation}
 {d\sigma\over d \alpha d^2p_{BT} d\theta_{cm}}=
  f(\alpha, p_{BT}) \phi(s', \theta_{cm}),
  \label{factoriz}
 \end{equation}
where $\alpha$ and $p_{BT}$ are the light-cone fraction 
and transverse momentum carried by $B$
(in the GPD notation $\alpha=(1- \xi)/(1+\xi)$, see section 3). 
The cross section is separated into the function $f$ (and $\phi$)  
which depends on the variables $\alpha$ and $p_{BT}$         
($s'$ and $\theta_{cm}$).                                      
The invariant energy of the subprocess is $s'=(1-\alpha)s$, and
the function $\phi$ is the cross section for the $q\bar q$-$3q$
scattering, and it is given by the quark counting rules 
as $\phi(s', \theta_{cm}) \approx \left( s' \right) ^n \gamma(\theta_{cm})$,
where $\gamma(\theta_{cm})$ is a function which depends only 
on the angle $\theta_{cm}$, and $n$ is the total number
of all interacting elementary fields. 
Phenomenologically, it appears more natural to use 
for numerical studies of experimental $s',t'$ dependence of
the elementary reaction $\pi N \to M N$, where $M$ is a meson, 
rather than the asymptotic quark counting  values.
 
Note the cross section as given by Eq. (\ref{factoriz}) does not 
depend on the direction of the transverse component of the momentum 
of $B$, or equivalently the cross section does not depend on
the angle between the plane formed by $B$ and the target nucleon 
and the plane of the hard reaction.
This provides an important test of factorization as
the rescatterings, which are discussed below, lead 
to break down of this factorization property.
 
\begin{figure}[t]
\begin{center}
\includegraphics[width=0.38\textwidth]{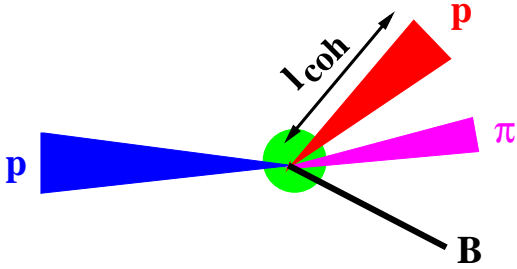}
\caption{Time evolution of wave packets of interacting hadrons
         in the $2\to3$ process.}
\label{fig:fmark}
\end{center}
\end{figure}

To determine at what $t'$ we may expect the onset of factorization,
we notice that the minimal condition is that $qqq$ and
$q\bar q$ wave packets, which are small in the interaction point,
involved in the hard interaction block should remain small
while passing by the spectator system. 
Hence, the distance over which a small wave package expands or contracts
- $ l_{coh}$,  should satisfy the condition
\begin{equation}
  l_{coh} > r_N \sim 0.8 \ {\rm fm}, 
\label{lcoh}
\end{equation}
corresponding to expansion of the outgoing hadrons and collapse
of incoming proton outside the target proton
as shown in Fig. \ref{fig:fmark}.
Here, 0.8 fm indicates the size of the nucleon.
A similar condition exists in the discussion of the color transparency
phenomena with nuclei where it is necessary to have 
$l_{coh} \ge 2 R_A$,
where $R_A$ is a nuclear radius, 
to reach the regime of complete color transparency. 
The studies of the evolution rates of $qqq$ and $q\bar q$ 
wave packets in the framework of the nuclear color transparency
have led to the estimates corresponding to 
(for the review of different models of space-time 
evolution of wave packages, see Ref. \cite{Frankfurt:1994hf})
 \begin{equation}
 l_{coh}= (0.3 \div 0.4 \ \mbox{fm}) \cdot p_h/({\rm GeV}/c).
 \end{equation}
In the case of pions, it is checked to a large extent
by the measurement of the A and $Q^2$ dependence of 
transparency in the $A(e,e'\pi)$ reaction \cite{:2007gqa}.
The condition 
of Eq. (\ref{lcoh}) is obviously much less stringent than
for the case of CT phenomena with nuclei.
It corresponds to requirement that final momenta of the two hadrons
with large $p_T$ are 
\begin{equation}
 p_c\ge 3 \div 4 \ {\rm GeV}/c, \ \ p_d\ge 3 \div 4 \ {\rm GeV}/c,
\end{equation}
which is satisfied for the incident nucleons 
with $p_N \ge 8$ GeV/$c$.

Spins and parities of $N$ and $\Delta$ are
$\frac{1}{2}^+$ and $\frac{3}{2}^+$, so that the state $h$
should be in $1^+$.
If it is a $p$-wave pion- or rho-like state, it is allowed. 
In general, $h$ could be expressed by the Fock-space expansion,
$|h>=|q\bar q>+|qq\bar q\bar q>+\cdot\cdot\cdot$. However, higher Fock
states are suppressed because more hard gluon exchanges are needed
to form the final pion and nucleon with large momenta. Therefore,
the intermediate state should be considered as a $q\bar q$ state.
Gluonic states such as $gg$ are excluded as the intermediate
state $h$ for $N \rightarrow \Delta$ because the isospin state 
of $h$ should be one by considering isospins of $N$ and $\Delta$ are 
$\frac{1}{2}$ and $\frac{3}{2}$, respectively. They are present 
for the $B=N$, though the data indicate \cite{White94} that
contribution of diagrams with gluon exchanges is much smaller
than that with quark exchanges.

If the final baryon ($B$) is a nucleon, there are additional
contributions in principle. The same discussions on isospin, 
spin, and parity suggest that the intermediate state should
be in $0^+$ and $1^+$ states. The $1^+$ state is the same
as the $N \rightarrow \Delta$ case.
There is an additional intermediate state with the vacuum quantum 
number $0^+$. Therefore, the Fock-space expansion is 
$|h>=|gg>+|q\bar q>+|qq\bar q\bar q>+$$\cdot\cdot\cdot$.
We do not take into account the $0^+$ intermediate state
in our current work with the following reasons. Lattice QCD
calculations indicate that the mass of the lightest $0^+$ glueball
is estimated about 1700 MeV \cite{lattice-glue}. 
Even if they were included in the cross-section estimate, 
their contributions are not significant in the kinematical
region of $|t| \ll m_N^2$ because of their large masses.
The $f_0 (600)$ (or $\sigma$) and $f_0 (980)$ mesons are not
$S$-wave $q\bar q$ states, so that their contributions are suppressed
in the limit $t' \to \infty$ with $t'/s'$=const as follows.
We have a meson-nucleon scattering amplitude which is proportional
to the integral of the initial and final meson light-cone waves function
over quark momenta: $\int \phi_M(z,\vec p_\perp) dz d^2p_\perp$,
where $z$ and $\vec p_\perp$ are the lightcone momentum fraction 
and transverse momentum of a quark, respectively.
However, wave functions of mesons which are in the $P$-wave
in the nonrelativistic limit have the property
$\phi_M(1-z,\vec p_\perp)= - \phi_M(z,\vec p_\perp)$
\cite{lightcone-wave}, so that the integral vanishes since the final-state
meson wave function is symmetric.
If the $f_0$ mesons are tetra-quark ($qq\bar q \bar q$) or meson-molecule
states \cite{f0}, their contributions are suppressed due to more
hard gluon exchanges.
In any case, it is possible to take channels, such as 
$p+p \rightarrow p+\pi^+ +n$, where the vacuum-like exchange
does not contribute. In this way, we consider only the $q\bar q$ state
as the intermediate state:
\begin{equation}
h = q\bar q .
\end{equation}

We would like to emphasize here that the practical problem we face
is that though it is possible to write the amplitude as a sum of
the terms representing convolutions of nucleon (transitional) GPDs
with the hard blocks $H$ of the $q\bar q$-$3q$ scattering,
these blocks are not known so far and hence generic factorized
expression is not useful for numerical calculations. 
Accordingly, we use an observation from the current studies of GPDs
that a distribution in the ERBL region over $z$
is to a very good accuracy symmetric 
and very close to that in the wave functions of $\pi$ and $\rho$ 
mesons which are close to the asymptotic form starting with
virtualities of a few GeV$^2$ \cite{gpv01}.
This will allow us to use the information about elementary
two body reactions. 

We explain our description of the nucleon and $N \rightarrow \Delta$
transition amplitudes ${\mathscr M}_{N\ra h B}$ in terms of the nucleonic
and transition GPDs in Sec. \ref{n-b-amplitude}, and the amplitude
${\mathscr M}_{h N\ra \pi N}$ is explained in Sec. \ref{piN}.


\section{Generalized parton distributions for nucleon 
         and $N \rightarrow \Delta$ transition}
\label{gpds}


\subsection{Generalized parton distributions for nucleon
            and $N \rightarrow \Delta$ transition}
\label{gpds-intro}

Generalized parton distributions are introduced in the $N \rightarrow B$
amplitude for the nucleon and $N \rightarrow \Delta$ for transitions 
with neutral exchange in $t$-channel ($p\to p$, $p\to \Delta^+$)
with other transitions related to these ones by the isospin symmetry.
The $N+N \rightarrow N + \pi + B$ cross section is factorized into
two amplitude in Eq. (\ref{eqn:factorization}). In this section, the first
term ${\mathscr M}_{N \rightarrow h B}$ is explained. 
The intermediate state $h$ should be considered as a $q\bar q$ state
for a reaction with large-momentum transfer $|t'|,\ |u'| \gg m_N^2$.
It is important to note that extraction of
such a $q \bar q$ state with a light-cone separation is related to
the GPDs in the ERBL kinematical region \cite{diehl03,erbl}. Therefore,
the amplitude ${\mathscr M}_{N \rightarrow h B}$ is simply given by
${\cal M}_{N \rightarrow B}$ times the factor describing 
a light-like separated $q\bar q$ pair.

The GPDs for the nucleon is given by off-forward matrix elements
of quark and gluon operators with a light-cone separation 
between nucleonic states \cite{diehl03,gpv01}.
The matrix element of vector current is associated with unpolarized GPDs
for quarks in the nucleon:
\begin{align}
 &  
 {\cal M}_{N}^{V} =
 \int\frac{d\lambda}{2\pi}e^{i\lambda x}
 \left< N, p_e \left| \overline{\psi}(-\lambda n/2) 
 \slashes{n} \psi(\lambda n/2) \right|N, p_a \right>
\nonumber \\
 & =
 I_N \, \overline{\psi}_N (p_e) 
 \big[ H (x,\xi,t) \slash{n} 
     + E (x,\xi,t)  \frac{i \sigma^{\alpha\beta} n_\alpha \Delta_\beta}{2m_N}
 \big ] \psi_N (p_a) ,
\label{eqn:unpol-n}
\end{align}
where 
$I_N$ is the isospin factor for the nucleon given by
$I_N = < 1/2 || \widetilde T || 1/2>
 \big < \frac{1}{2} M_N : 1 m \big | \frac{1}{2} M_N^\prime \big > /\sqrt{2}$
\cite{n-delta-isospin}:
\begin{equation}
I_N = 1, \ \sqrt{2} \ \ \text{for $p \rightarrow p$, $n$},
\end{equation}
$\sigma^{\alpha\beta}$ is defined by 
$\sigma^{\alpha\beta}=(i/2)[\gamma^\alpha, \gamma^\beta]$,
and $n^{\mu}$ is a light-cone null vector which satisfies
\begin{equation}
n^2 = 0, \ \  n\cdot P=1 
\ \ \text{for }  P=\frac{1}{2}(p_a + p_e) .
\end{equation}
The vector $n^\mu$ is chosen so as to have $n^+ = \vec n_\perp=0$,
where $\vec n_\perp$ is the transverse vector and $n^\pm$ is defined by
$n^\pm = (n^0 \pm n^3)/\sqrt{2}$. Then, $n_-^{\mu}$ is simply denoted
as $n^\mu$ in the following discussions \cite{difference,note-n}:
\begin{equation}
n^\mu \equiv n_-^{\mu} = 
  \left ( \frac{1}{2 P^+}, \ 0, \ 0, \ - \frac{1}{2 P^+} \right ) .
\end{equation}
The momentum $P$ is the average over the initial and final
nucleon momenta. The integral variable $\lambda$ is defined by
$\lambda = P \cdot z \simeq P^+ z^-$.
In Eq. (\ref{eqn:unpol-n}), $\psi(\lambda n/2)$ is the quark field.
The functions $H (x,\xi,t)$ and $E (x,\xi,t)$ are the unpolarized
GPDs for the nucleon.
The Dirac spinor for the nucleon is denoted as $\psi_N (p)$.
Quark-spin dependent GPDs are also defined in the similar way
for the nucleon:
\begin{align}
 & \! \! \! \! \! \! 
 {\cal M}_{N}^{A} =
 \int\frac{d\lambda}{2\pi}e^{i\lambda x}
 \left< N, p_e \left| \overline{\psi}(-\lambda n/2) 
 \slashes{n} \gamma_5 \psi(\lambda n/2) \right|N, p_a \right>
\nonumber \\
 & =
 I_N \, \overline{\psi}_N (p_e) 
 \big[ \widetilde H (x,\xi,t) \slash{n} \gamma_5
     + \widetilde E (x,\xi,t)  \frac{n \cdot \Delta \gamma_5}{2m_N}
 \big ] \psi_N (p_a) ,
\label{eqn:pol-n}
\end{align}
where $\widetilde H (x,\xi,t)$ and $\widetilde E (x,\xi,t)$ are
the polarized GPDs for the nucleon.

\begin{figure}[t]
\begin{center}
\includegraphics[width=0.38\textwidth]{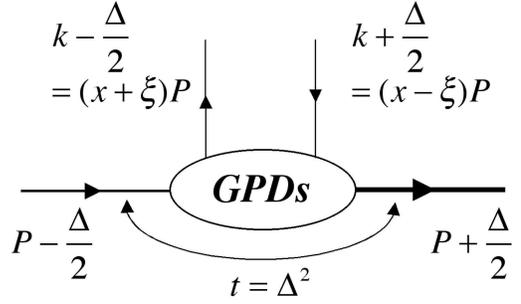}
\caption{Kinematics for the $N \rightarrow B$ transition
 and generalized parton distributions.}
\label{fig:gpd}
\end{center}
\end{figure}

The generalized scaling variable $x$ and a skewdness parameter $\xi$
are defined in terms of light-cone momentum $k^+$ ($= (k^0 +k^3)/\sqrt{2}$)
and $P^+$ as
\begin{equation}
k^+ =x P^+ , \ \  \Delta^+ 
= -2\xi P^+, \ \  \xi=-\frac{1}{2} n\cdot\Delta , 
\label{eqn:xi}
\end{equation}
where $k$ is the average momentum of the quarks,
$\Delta^{\mu}$ is the four momentum transfer from the initial nucleon
$(a)$ to the final nucleon $(e)$ (or $\Delta$ in the $N \rightarrow \Delta$
transition), and $t$ is its squared quantity:
\begin{equation}
\Delta = p_e - p_a , \ \  t\equiv(p_e - p_a)^2=\Delta^2 .
\end{equation}
A schematic diagram is illustrated in Fig. \ref{fig:gpd} for understanding
the kinematics of the GPDs.
The variable $x$ indicates the momentum fraction carried by a quark
in the nucleon as usually used in parton distribution functions (PDFs).
The skewdness parameter $\xi$ or the momentum $\Delta$ indicates
the momentum transfer from the initial nucleon to the final one
or the one between the quarks. 
In the forward limit ($\Delta \rightarrow 0$), the nucleonic GPDs
$H (x,\xi,t)$ and $\widetilde H (x,\xi,t)$ become usual PDFs 
for the nucleon as shown in Sec. \ref{n-gpds-prop}.

\begin{figure*}[t]
\begin{center}
\includegraphics[width=0.80\textwidth]{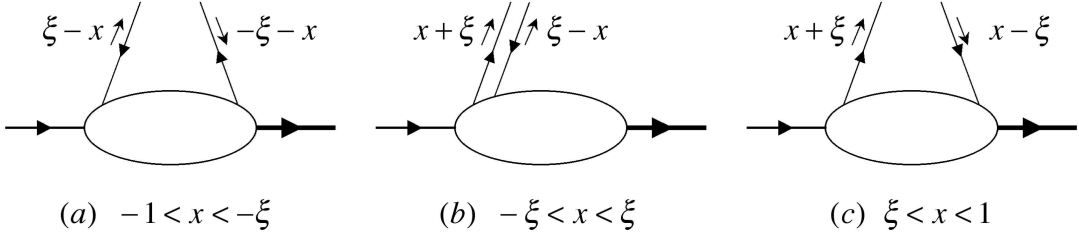}
\caption{Three regions of $x$ for the GPDs:
(a) $-1 < x < -\xi$, 
    emission and reabsorption of an antiquark,
(b) $-\xi < x < \xi$, 
    emissions of a quark and an antiquark,
(c) $\xi < x < 1$, 
    emission and reabsorption of a quark.}
\label{fig:3gpd-regions}
\end{center}
\end{figure*}

Next, we explain how to express the transition
amplitude ${\cal M}_{N \rightarrow \Delta}$ in terms of the transition
GPDs for $N \rightarrow \Delta$.
First, the $N \rightarrow \Delta$ spin-independent quark
distributions are defined by a product of quark fields 
at a light-cone separation between nucleon and $\Delta$ states
\cite{gpv01}:
\begin{align}
 & \! \! \! \! \! \! 
 {\cal M}_{N \ra \Delta}^{V} =
 \int\frac{d\lambda}{2\pi}e^{i\lambda x}
 \left< \Delta, p_e \left| \overline{\psi}(-\lambda n/2) 
 \slashes{n} \psi(\lambda n/2) \right|N, p_a \right>
\nonumber \\
 & = I_{\Delta N}
 \overline{\psi}_{\Delta}^{\, \mu}(p_e) 
 \big[ H_M (x,\xi,t){\cal K}_{\mu\nu}^M n^{\nu} 
     + H_E (x,\xi,t){\cal K}_{\mu\nu}^E n^{\nu} 
\nonumber \\
 &  \ \ \ \ \ \ \ \ \ \ \ \ \ \ \ \ \ \ \ 
     + H_C (x,\xi,t){\cal K}_{\mu\nu}^C n^{\nu}  \big] \psi_N (p_a) ,
\label{eqn:unpol-n-delta}
\end{align}
where $I_{\Delta N}$ is the $N \rightarrow \Delta$ transition
isospin factor.
The terms ${\cal K}_{\mu\nu}^M$, ${\cal K}_{\mu\nu}^E$, 
and ${\cal K}_{\mu\nu}^C$ indicate covariants associated with
magnetic dipole, electric quadrupole, and Coulomb quadrupole 
transitions \cite{js73}:
\begin{align}
{\cal K}_{\mu\nu}^M
& = -i\frac{3(m_\Delta+m_N)}{2m_N[(m_\Delta+m_N)^2-t]}
\varepsilon_{\mu\nu\lambda\sigma} P^\lambda
\Delta^\sigma ,
\nonumber \\
{\cal K}_{\mu\nu}^E
& = -{\cal K}_{\mu\nu}^M 
-\frac{6(m_\Delta+m_N)}{m_NZ(t)}
\varepsilon_{\mu\sigma\lambda\rho} P^\lambda\Delta^\rho
\varepsilon^\sigma_{\nu \kappa\delta} P^\kappa\Delta^\delta
\gamma^5, 
\nonumber \\
{\cal K}_{\mu\nu}^C
& = -i\frac{3(m_\Delta+m_N)}{m_NZ(t)} \Delta_\mu
(t P_\nu-\Delta\cdot P\Delta_\nu)\gamma^5 ,
\label{eqn:K}
\end{align}
where $m_\Delta$ is the $\Delta$ mass, and $Z(t)$ is defined by
\begin{equation}
Z(t)=[(m_\Delta+m_N)^2-t][(m_\Delta-m_N)^2-t].
\end{equation}
Our conventions for the antisymmetric tensor
$\varepsilon_{\beta\mu\lambda\sigma}$ and $\gamma^5$ are taken as
$\varepsilon_{0123}=+1$ and 
$\gamma^5=i \gamma_0 \gamma_1 \gamma_2 \gamma_3$
\cite{gpv01,diehl03,bd-book}.
The Rarita-Schwinger spinor \cite{Rarita-Schwinger} for the $\Delta$
is denoted as ${\psi}_{\Delta}^{\mu}(p)$.

The isospin factor $I_{\Delta N}$ in Eq. (\ref{eqn:unpol-n-delta}) comes
from the transition isospin. The $N \rightarrow \Delta$ transition isospin
is define by \cite{n-delta-isospin}
\begin{align}
\bigg < \frac{3}{2} M_\Delta \bigg | \, \widetilde T \,
\bigg | \frac{1}{2} M_N \bigg >
= & \sum_m \frac{1}{2} \bigg < \frac{3}{2} \bigg | \bigg | \, \widetilde T \,
\bigg | \bigg | \frac{1}{2} \bigg >
\nonumber \\
& \times \bigg < \frac{1}{2} M_N : 1 m \bigg | \frac{3}{2} M_\Delta \bigg > \,
\tilde\varepsilon_m ,
\end{align}
where $\tilde\varepsilon_m$ is the spherical unit vector,
$< 3/2 || \, \widetilde T \, | | 1/2 > =2$ is a reduced matrix element,
and $< 1/2 \, M_N : 1 \, m | 3/2 \, M_\Delta >$ is
a Clebsch-Gordan coefficient. The isospin factor becomes 
\begin{equation}
\! \! \! \!
I_{\Delta N} = 1, \ \sqrt{\frac{2}{3}}, \ \sqrt{\frac{1}{3}}, \ 0 \ \ 
\text{for $p \rightarrow \Delta^{++}, \Delta^+, \ \Delta^0, \ \Delta^-$}, 
\end{equation}
respectively.

In Eq. (\ref{eqn:unpol-n-delta}), the functions $H_M (x,\xi,t)$,
$H_E (x,\xi,t)$, and $H_C (x,\xi,t)$ are the unpolarized GPDs
for the transition $N \rightarrow \Delta$.
Quark-spin dependent GPDs are defined in the same way \cite{gpv01}:
\begin{align}
 & \! \! \! \! \! \! \! \! \! \! 
 {\cal M}_{N \ra \Delta}^{A} =
 \int\frac{d\lambda}{2\pi}e^{i\lambda x}
 \left< \Delta, p_e \left| \overline{\psi}(-\lambda n/2) 
 \slashes{n}\gamma^5 \psi(\lambda n/2) \right|N, p_a \right>
 \ \ \ \ \ 
\nonumber \\
 & =  I_{\Delta N} \,
 \overline{\psi}_{\Delta}^{\, \mu}(p_e) \bigg [ \widetilde{H}_1 (x,\xi,t) n_{\mu} 
   + \widetilde{H}_2 (x,\xi,t)\frac{\Delta_{\mu}(n\cdot\Delta)}{m_N^2} 
\nonumber \\
 & \ \ \ \ \ \ \ \ \ \ \ \ \ \ \ 
   + \widetilde{H}_3 (x,\xi,t)\frac{n_{\mu}\slash{\Delta}
                                -\Delta_{\mu}\slash{n}}{m_N} 
\nonumber \\
 & \ \ \ \ \ \ \ \ \ \ \ \ \ \ \ 
  + \widetilde{H}_4 (x,\xi,t) \frac{{P}\cdot\Delta n_{\mu}
 -2\Delta_{\mu}}{m_N^2} \bigg ] \psi_N (p_a), 
\label{eqn:pol-n-delta}
\end{align}
where the functions 
$\widetilde{H}_i (x,\xi,t)$ ($i$=1, 2, 3, or 4) \cite{h-notation}
are the polarized GPDs for the $N \rightarrow \Delta$ transition.

Partonic interpretation of the GPDs is different depending on $x$ regions
as illustrated in Fig. \ref{fig:3gpd-regions}. The $x$ variable is divided
into three regions, $-1 < x < -\xi$, $-\xi < x < \xi$, and $\xi < x < 1$,
and the GPDs are interpreted as follows in each $x$ region
\cite{diehl03}:
\begin{enumerate}
\item[(a)] ``Antiquark distribution" \\ 
            $-1 < x < -\xi$ \ ($x+\xi <0, x-\xi <0$) \\
 $\bullet$ Emission of antiquark with momentum fraction $\xi -x$ \\
 $\bullet$ Absorption of antiquark with momentum fraction $-x-\xi$  \\
 Both momentum fractions $x+\xi$ and $x-\xi$ are negative, so that 
 an emission (absorption) of a quark with the negative momentum fraction
 $x+\xi$ ($x-\xi$) corresponds to an absorption (emission) of an antiquark
 with the positive momentum fraction $-x-\xi$ ($-x+\xi$).
\item[(b)] ``Meson (quark-antiquark) distribution amplitude" \\
            $-\xi < x <\xi$ \ ($x+\xi >0, x-\xi <0$) \\
 $\bullet$ Emission of quark with momentum fraction $x+\xi$ \\
 $\bullet$ Emission of antiquark with momentum fraction $\xi -x$ \\
 The momentum fraction $x-\xi$ is negative, so that an absorption of 
 a quark with the negative momentum fraction $x-\xi$ corresponds to
 an emission of an antiquark with the positive momentum fraction $-x+\xi$.
\item[(c)] ``Quark distribution" \\
            $\xi < x <1$ \ ($x+\xi >0, x-\xi >0$) \\
 $\bullet$ Emission of quark with momentum fraction $x+\xi$ \\
 $\bullet$ Absorption of quark with momentum fraction $x-\xi$ \\
\end{enumerate}
The regions (a) and (c) are called as
DGLAP (Dokshitzer-Gribov-Lipatov-Altarelli-Parisi) regions, and
(b) is called the ERBL region \cite{erbl} depending on evolution types.

If the reaction $N \rightarrow h(q \bar q)+\Delta$ in Fig. \ref{fig:diagram2}
is dominated by the kinematical region $-\xi < x < \xi$, the cross section of
$p+p \rightarrow p+\pi+\Delta$ should be expressed by the transition GPDs
for $N \rightarrow \Delta$. They correspond to the distributions in
the kinematical region $(b)$ in Fig. \ref{fig:3gpd-regions}. 
The GPDs for the nucleon are calculated, for example, in a chiral soliton model,
and results indicated that the GPDs are dominated by a meson-pole
contribution at small momentum transfer and that the distributions
are located dominantly in the region $-\xi < x < \xi$ \cite{ppg00}.
This supports our treatment of the blok $N \rightarrow h(q \bar q)+\Delta$,
in the calculation of the process $N+N \rightarrow N+\pi+\Delta$,
in terms of the transition GPDs.

\subsection{Properties of generalized parton distributions for nucleon}
\label{n-gpds-prop}

Here we introduce basic properties of the GPDs for the nucleon.
In the forward limit ($\xi=t=0$), the GPDs $H$ and $\widetilde H$
for quarks are given by
\begin{equation}
           H (x, 0, 0) = q(x), \ \ \ 
\widetilde H (x, 0, 0) = \Delta q(x) .
\end{equation}
where $q(x)$ and $\Delta q(x)$ are unpolarized and polarized
parton distribution functions in the nucleon.
Next, their first moments are the form factors of the nucleon:
\begin{alignat}{2}
\int_{-1}^{1} dx            H(x,\xi,t) & = F_1 (t), \  &
\int_{-1}^{1} dx            E(x,\xi,t) & = F_2 (t),
\nonumber  \\
\int_{-1}^{1} dx \widetilde H(x,\xi,t) & = g_A (t), \  &
\int_{-1}^{1} dx \widetilde E(x,\xi,t) & = g_P (t),
\end{alignat}
where $F_1 (t)$, $F_2 (t)$, $g_A (t)$, and $g_P (t)$ are Dirac, Pauli,
axial, and pseudoscalar form factors of the nucleon.
A second moment gives a quark orbital-angular-momentum 
contribution ($L_q$) to the nucleon spin:
\begin{align}
   J_q & = \frac{1}{2} \int dx \, x \, [ H (x,\xi,t=0) +E (x,\xi,t=0) ]
\nonumber \\
       & = \frac{1}{2} \Delta q + L_q ,
\label{eqn:Jq}
\end{align}
where $\Delta q$ is the quark-spin contribution and $J_q$ is
the total angular-momentum of quarks. It indicates that
nucleonic GPD studies are important for clarifying the origin
of the nucleon spin. As mentioned in the introduction,
the quark contribution to the nucleon spin, possibly as well as
the gluon one, is small, so that orbital angular momenta play
an important role as the origin of the nucleon spin.
These facts indicate that the GPDs are important for understanding
basic properties of the nucleon from low to high energies.

\subsection{Properties of generalized parton distributions 
            for $N \rightarrow \Delta$ transition}
\label{n-delta-gpds-prop}

In this section, basic properties of the $N \rightarrow \Delta$
transition GPDs are explained. 
First, the first moment of the GPDs corresponds to
the $N \rightarrow \Delta$ transition form factors
\cite{gpv01,diehl03}:
\begin{equation}
\int_{-1}^1 dx H_{i}(x,\xi,t)=2G_{i}(t) , \ \ i=M,\ E,\ C 
\end{equation}
where $G_M (t)$, $G_E (t)$, and $G_C (t)$ are the magnetic dipole,
electric quadrupole, and Coulomb quadrupole transition form factors,
respectively. The $N \rightarrow \Delta$ transition is dominated
by the M1 transition; however, there are also small E2 and C2
contributions \cite{NDelta-form,NDelta-E2}. 
Experimental investigations have been done in electron scattering, 
particularly in the processes such as $(e,e' \pi)$ 
\cite{NDelta-form,NDelta-E2,ujs07}.
The E2 and C2 transitions are especially important for investigating
a dynamical aspect of baryon structure because they are related to
the deformation of the nucleon and $\Delta$.
It is interesting to note that information on such a deformation
should be also contained in the GPDs. The studies of
the $N \rightarrow \Delta$ transition GPDs should be valuable
for such a new aspect of low-energy hadron structure.

The first moments of $\widetilde H_i (x,\xi,t)$ ($i$=1, 2, 3, 4)
are axial form factors for the $N \rightarrow \Delta$ transition:
\begin{alignat}{2}
& \! \! 
\int_{-1}^1 dx \widetilde H_1 (x,\xi,t) = 2 \, C_5^A(t), \ \ 
& \! 
\int_{-1}^1 dx \widetilde H_2 (x,\xi,t) = 2 \, C_6^A(t), &
\nonumber \\
& \! \! 
\int_{-1}^1 dx \widetilde H_3 (x,\xi,t) = 2 \, C_3^A(t), \ \ 
& \! 
\int_{-1}^1 dx \widetilde H_4 (x,\xi,t) = 2 \, C_4^A(t), &
\label{eqn:1st-moment-pol}
\end{alignat}
where $C_i^A(t)$ ($i$=3, 4, 5, or 6) are the axial form factors
in Ref. \cite{adler75}. These axial form factors were studied 
in parity-violating electron scattering \cite{parity-electron}.
These Adler form factors are estimated in various theoretical
models \cite{parity-electron}.
For example, a recent estimation by the chiral perturbation theory
indicates that $C_3^A$ and $C_4^A$ vanish in the leading order
and they become finite due to higher-order corrections:
$ | C_3^A(t) |, | C_4^A(t) | \ll 1$ \cite{geng-etal-08}.

Next, the $N \rightarrow \Delta$ GPD studies are also valuable for
understanding the origin of the nucleon spin, especially
orbital-angular-momentum effects.
As an example, possible connections to the nucleon spin are shown
by taking the large $N_c$ limit. In the chiral-soliton model,
the $N\ra\Delta$ transition GPDs are related to the nucleonic GPDs
because both baryons are different rotational states of the same soliton
\cite{gpv01}: 
\begin{align}
H_M(x,\xi,t) = & 
    \frac{2}{\sqrt{3}} \left[ E^{u}(x,\xi,t)-E^{d}(x,\xi,t) \right] ,
\nonumber \\
\widetilde{H}_1(x,\xi,t) =&
    \sqrt{3} \left[ \widetilde{H}^{u}(x,\xi,t)-\widetilde{H}^{d}(x,\xi,t) \right] ,
\nonumber \\
\widetilde{H}_2(x,\xi,t) =&
    \frac{\sqrt{3}}{4}\left[ \widetilde{E}^{u}(x,\xi,t)
                            -\widetilde{E}^{d}(x,\xi,t) \right] ,
\nonumber \\
H_E(x,\xi,t) =& H_C(x,\xi,t)=\widetilde{H}_3(x,\xi,t)
\nonumber \\
             =& \widetilde{H}_4(x,\xi,t)=0 ,
\label{eqn:h-large-nc}
\end{align}
in the leading order of $1/N_c$ expansion.
It is especially noteworthy that the second moment of $H_M(x,\xi,t)$
is related to the isovector part of the angular momentum of the nucleon
carried by quarks as
\begin{align}
\lim_{t\ra 0, N_c \ra\infty} & \int_{-1}^1 dx\ xH_{M}(x,\xi,t)
\nonumber \\
& = \frac{2}{\sqrt{3}} \left[ 2(J^u -J^d)-M_2^u +M_2^d \right] , 
\label{eqn:Ju-d}
\end{align}
where 
\begin{equation}
M_2^q = \int_0^1 dx \ x\left[ q(x)+\bar{q}(x) \right] , 
\end{equation}
is a second moment of quark and antiquark distribution.
The $J^u$ and $J^d$ indicate total angular momenta carried by $u$ and
$d$ quarks, respectively. They are equal to the sum of 
the spin and orbital angular momentum carried by the quark,
as shown in Eq. (\ref{eqn:Jq}).
Equation (\ref{eqn:Ju-d}) suggests that the $N \rightarrow \Delta$ GPDs
should be important for investigating the isovector part the orbital
angular momentum, $L^u-L^d$, in the studies of nucleon spin.

\section{$N\rightarrow B$ transitions
          in terms of generalized parton distributions}
\label{n-b-amplitude}

\subsection{Generalized parton distributions for the nucleon}
\label{n-n-cross}

For calculating the unpolarized cross section of 
$N+N \rightarrow N+\pi+N$, spin summations are taken for the nucleons.
In this section, the factor describing transition 
from the initial nucleon $(a)$ to the final one $(e)$ is calculated.
We discussed the amplitudes $M_N^V$ and $M_N^A$, which are described 
by spin independent and dependent generalized quark distributions. 
The absolute-value squared of the transition amplitude is given by
\begin{equation}
\sum_{\lambda_{N}, \lambda_{N'}} |{\cal M}_{N}|^2 
= \sum_{\lambda_{N}, \lambda_{N'}} 
\left( |{\cal M}_{N}^{V}|^2 + |{\cal M}_{N}^{A}|^2 \right) . 
\label{eqn:n-emplitude2}
\end{equation}
It should be noted that there is no interference term
if the spin summations are taken over $\lambda_N$ ($=\lambda_a$) 
and $\lambda_{N'}$ ($=\lambda_e$).
The summation of the vector and axial contributions indicate that
all the associated $q\bar q$ states should be included in calculating
the cross section. The calculations are done in the same way as
mesonic contributions to antiquark distributions
\cite{pi-n-delta-calc}.

Using the spin summation of the Dirac spinors,
$ \sum_{\lambda_N}\psi_N (p)\overline{\psi}_N (p) = \slashes{p} -m_N $
and Eq. (\ref{eqn:unpol-n}) for ${\cal M}_{N}^{V}$, 
and then calculating traces,
we express the vector part of Eq. (\ref{eqn:n-emplitude2})
in terms of nucleonic GPDs:
\begin{align}
& \! \! \! \! 
\sum_{\lambda_{N}, \lambda_{N'}} |{\cal M}_{N}^{V}|^2
 = I_{N}^{\ 2} \bigg [ 8 (1-\xi^2) \{ H(x,\xi,t) \}^2
\nonumber \\
& \! \! \! \! 
+ 16 \xi^2 H(x,\xi,t) E(x,\xi,t) 
- \frac{t}{m_N^2} (1+\xi)^2 \{ E(x,\xi,t) \}^2 \bigg ] .
\label{eqn:MV-N}
\end{align}
In the same way, the axial term is calculated as
\begin{align}
& \! \! \! \!
\sum_{\lambda_{N}, \lambda_{N'}} |{\cal M}_{N}^{A}|^2
 = I_{N}^{\ 2} 
\bigg [ 8 (1-\xi^2) \{ \widetilde H(x,\xi,t) \}^2
\nonumber \\  
& \ \ \                  
+ 18 \xi^2 \widetilde H(x,\xi,t) \widetilde E(x,\xi,t) 
- \frac{2 \, t \, \xi^2}{m_N^2}
  \{ \widetilde E(x,\xi,t) \}^2 \bigg ] .
\label{eqn:MA-N}
\end{align}
These expressions are used for calculating the cross section
for $N+N \rightarrow N+\pi+N$.

\subsection{Generalized parton distributions for the
            $\mathbf {N \rightarrow\Delta}$ transition}
\label{n-delta-cross}

The $N \rightarrow \Delta$ transition part is calculated
in the same way by taking spin summations for the nucleon
and the $\Delta$. The $N\ra\Delta$ transition was described by
the matrix elements ${\cal M}_{N\ra \Delta}^{V}$ and 
${\cal M}_{N\ra \Delta}^{A}$ with the transition GPDs
in Sec. \ref{gpds-intro}.
Then, absolute-value squared of the transition amplitude is
\begin{equation}
\! \! \! \! 
\sum_{\lambda_N, \lambda_\Delta} |{\cal M}_{N \ra \Delta}|^2 
= \sum_{\lambda_N, \lambda_\Delta} 
\left( |{\cal M}_{N \ra \Delta}^{V}|^2 
     + |{\cal M}_{N \ra \Delta}^{A}|^2 \right) . 
\label{eqn:n-delta-emplitude2}
\end{equation}
Because the spin summations are taken over $\lambda_N$ and 
$\lambda_\Delta$, there is no interference term.

In our numerical estimation for the cross section, we use 
the following approximations.
First, the electric quadrupole contributions $H_E$ and $H_C$ should
be small as suggested in theoretical and experimental studies in
the $N \rightarrow \Delta$ transition form factors 
\cite{NDelta-form, NDelta-E2}. Therefore, it is appropriate
to assume $H_E(x,\xi,t)$=$H_C(x,\xi,t)$=0 for our first estimation
of the cross section if such small effects can be neglected.
This is also satisfied in the chiral-soliton model as shown
in Eq. (\ref{eqn:h-large-nc}). Then, the vector amplitude
is expressed only by the magnetic term:
\begin{equation}
{\cal M}_{N\ra \Delta}^{V} 
= I_{N\Delta} \overline{\psi}_{\Delta}^{\, \mu}(p_e)
 H_M (x,\xi,t){\cal K}_{\mu\nu}^M n^{\nu} \psi_N (p_a) .
\label{eqn:mv}
\end{equation}
Taking the square of its absolute value and 
performing spin summations, we obtain
\begin{align}
& 
\sum_{\lambda_N, \lambda_\Delta} |{\cal M}_{N \ra \Delta}^{V}|^2
 = - I_{N\Delta}^{\ 2} \left\{ H_M (x,\xi,t) \right\} ^2
\nonumber \\  
& \! \! \! 
\times\, {\rm Tr}\left[ \sum_{\lambda_\Delta}
\psi_{\Delta}^{\mu}(p_e)\overline{\psi}_{\Delta}^{\, \alpha}(p_e) 
{\cal K}_{\mu\nu}^M n^{\nu} \sum_{\lambda_N}\psi_N (p_a)\overline{\psi}_N (p_a) 
{\cal K}_{\alpha\beta}^M n^{\beta} \right] . 
\label{MEofUNPOL}
\end{align}
Spin summations of the Rarita-Schwinger spinors are given by
\cite{Rarita-Schwinger}
\begin{align}
& \sum_{\lambda_\Delta}\psi_{\Delta}^{\mu} (p_e)
\overline{\psi}_{\Delta}^{\, \alpha} (p_e) = (\slash{p}_e -m_{\Delta})
\nonumber \\
& \ \ 
\times \left( -g^{\mu\alpha} 
+\frac{1}{3}\gamma^{\mu}\gamma^{\alpha} 
-\frac{p_e^{\mu}\gamma^{\alpha}-p_e^{\alpha}\gamma^{\mu}}{3m_{\Delta}} 
+\frac{2p_e^{\mu}p_e^{\alpha}}{3m_{\Delta}^2} \right) ,
\label{eqn:rarita}
\end{align}
where $m_\Delta$ is the $\Delta$ mass.
We denote the trace part of Eq. (\ref{MEofUNPOL}) as $C_M (\xi, t)$:
\begin{align}
& \sum_{\lambda_N, \lambda_\Delta} |{\cal M}_{N \ra \Delta}^{V}|^2
=  I_{N\Delta}^{\ 2}  C_M (\xi, t) \{ H_M (x,\xi,t) \}^2 ,
\label{eqn:mv-n-d}
\end{align}
and it is calculated by using Eq. (\ref{eqn:rarita}) as
\begin{align}
& C_M (\xi, t) = 3 \, 
   \left [ \frac{ m_N+m_{\Delta} }
             {m_N \{ (m_N+m_{\Delta})^2-t \} }
   \right ] ^2 
\nonumber \\ 
& 
\times  \left[ t-(m_N+m_{\Delta})^2 \right ] 
\nonumber \\ 
& 
\times
\left[ t (1 - \xi^2)+ 2\xi (m_{\Delta}^2 -m_N^2) 
       + 2\xi^2 (m_{\Delta}^2 +m_N^2) \right] . 
\label{CFofUNPOL}
\end{align}
\vspace{0.2cm}

Similarly, the axial-vector part is given by (neglecting the GPDs 
$H_3(x,\xi,t)$ and $H_4(x,\xi,t)$ as suggested, for example, by 
the chiral-soliton model \cite{gpv01, fpsv9800} and also by 
the chiral perturbation theory of the form factors \cite{geng-etal-08}):
\begin{align}
{\cal M}_{N\ra \Delta}^{A} 
=  I_{N\Delta} & \, \overline{\psi}_{\Delta}^{\, \mu}(p_e) 
\bigg [ \tilde{H}_1 (x,\xi,t) n_{\mu} 
\nonumber \\
& + \tilde{H}_2 (x,\xi,t)\frac{\Delta_{\mu}(n\cdot\Delta)}{m_N^2}
 \bigg ] \psi_N (p_a).
\label{eqn:ma}
\end{align}
Expressing these amplitudes in terms of the GPDs in Eqs. (\ref{eqn:mv})
and (\ref{eqn:ma}), we obtain
\begin{align}
\sum_{\lambda_N, \lambda_\Delta} & |{\cal M}_{N\ra \Delta}^{A}|^2 
=  I_{N\Delta}^{\ 2} \left [
C_1 (\xi,t) \{ \widetilde{H}_1 (x,\xi,t) \}^2   \right.
\nonumber \\ 
&  
+  C_2 (\xi,t) \{ \widetilde{H}_2 (x,\xi,t) \}^2 
\nonumber \\
& 
\left.
+ C_{12} (\xi,t)  \widetilde{H}_1 (x,\xi,t) 
                  \widetilde{H}_2 (x,\xi,t) \right ] ,
\label{MEofPOL}
\end{align}
where $C_{i}(\xi,t)$ are coefficient terms and their explicit forms are
\begin{align}
C_{1}(\xi,t) & =  \frac{4}{3m_{\Delta}^2} \, (1-\xi)^{2} \,
                  \big [ (m_{\Delta} + m_N)^2 -t \big ] , 
\label{CFofPOL1}
\\
C_{2}(\xi,t) & =  \frac{4}{3 m_N^4 m_{\Delta}^2} \, \xi^2 \,
                  \big [ (m_{\Delta} + m_N)^2 -t \big ]
\nonumber \\
   & \ \ \  \ \ \ 
        \times    \big [ -4 m_\Delta^2 t + ( m_\Delta^2 -m_N^2+t)^2 \big ] ,
\label{CFofPOL2}
\\
C_{12}(\xi,t) & =  -  \frac{8 \xi}{3 m_N^2 m_{\Delta}^2} \,
                     \big [ (m_{\Delta} + m_N)^2 -t \big ] \,
\nonumber \\
   & \ \ \  \ \ \ 
        \times \big [ 4 m_\Delta^2 \xi + (1-\xi)( m_\Delta^2 -m_N^2 +t) \big ] .
\label{CFofPOL3}
\end{align}

\noindent
In this way, the $N \rightarrow \Delta$ transition part
of the cross section can be calculated using the transition GPDs. 

\section{$\mathbf{h N\ \rightarrow \pi N}$ scattering 
at large momentum transfer}
\label{piN}

\subsection{Intermediate $\mathbf{q\bar q}$ and mesons}
\label{qqbar-pion}

As explained in Sec.\ref{gpds-intro}, the GPDs are separated into
three $x$ regions. If the momentum transfer $|t|$ is small,
the GPDs distributed dominantly in the $-\xi<x<\xi$, for example,
as suggested in Ref. \cite{ppg00}. 
The GPDs in the kinematical region, $-\xi<x<\xi$, are the distribution
amplitudes with the quark and antiquark emissions with momentum
fractions $x+\xi$ and $-x+\xi$, respectively.

In general, all associated $q\bar q$ states should contribute to 
the cross section as we considered the vector- and axial-vector 
amplitudes. At  small $|t| \ll m_N^2$,
light-meson-like $q\bar q$ pairs contribute mostly 
to the cross section. The dominant ones are
the $q\bar q$ state with the pion ($\rho$) quantum numbers
in the axial-vector (vector) amplitude.
Then, the $q\bar q$ pair could be considered as a pion or $\rho$
state in a small-size configuration and the amplitude
is given by the meson ($M=\pi$ or $\rho$) wave function $\phi_M (z)$,
and its normalization is given by the Brodsky-Lepage relation \cite{bl,rho}.
However, our cross sections are independent of the functional form
of this wave function in the region where interpolation of the GPDs 
by the lowest states is a good approximation (see Sec. \ref{results}).
The kinematical variable $z$ is the light-cone momentum fraction
and it is defined in the following way.
The variables $x_1$ and $x_2$ indicate momentum fractions carried by
a quark and an antiquark, respectively, in the nucleon or
the $N \rightarrow \Delta$ transition. As indicated in Fig. \ref{fig:gpd}, 
they are expressed by $x$ and $\xi$ as
\begin{equation}
x_1= x+\xi, \ \ 
x_2=-x+\xi.
\end{equation}
The variable $z$ is the momentum fraction carried by the quark to 
the total momentum of the quark and antiquark system:
\begin{equation}
z=\frac{x_1}{x_1+x_2}=\frac{x+\xi}{2\xi} .
\end{equation}

In the previous section, the $N \rightarrow B$ part was calculated
in terms of the nucleonic  and $N \rightarrow \Delta$ transition GPDs,
which are associated mainly with the $q \bar q$ emission.
In the small momentum transfer region ($|t| \ll m_N^2$), 
the $q\bar q$ pair could be considered as a pion or $\rho$ state
in a small-size configuration, which next interacts with another proton
$(b)$ as shown in Fig. \ref{fig:diagram2}. The probability that the $q\bar q$
pair forms a meson $M$ is described by the meson wave function
in the light-cone coordinates. However, it should be noted that
the GPDs at the meson pole effectively contain the wave function of 
the meson in the point like configuration ($\phi_M (z)$) as indicated, 
for example, in Ref. \cite{gpv01} as the pion-pole contributions.
This factor enters also in the meson-nucleon scattering amplitude, 
so that this factor should be removed in matching of the GPD
description with description in terms of scattering off a meson:
\begin{equation}
{\mathscr M}_{(q\bar{q})_M p\ra \pi p}= {\cal M}_{M p\ra \pi p}
/\phi_M (z),
\end{equation}
where $(q\bar{q})_M$ is a $q\bar q$ state with the quantum numbers
of the meson $M$. The critical point here 
is the observation of several GPD analyses that $z$ distribution
of $q\bar q$ pairs in the GPDs is close to that in the pion
and rho mesons.
The off-shell nature of the $q\bar q$-$N$ scattering amplitude
can be best seen from 
$s'=-t'- u' +2 m_N^2 +m_{\pi}^2+t$ 
as compared to the on-shell case of 
$s'=-t'- u' +2 m_N^2 +m_{\pi}^2+m_M^2$,
where $m_M$ is the meson mass.
However, in the limit Eq. (\ref{asympt}), the cross section 
is a function of the ratio of large variables and so correction 
goes to zero. 

For the spin-dependent GPDs in the axial amplitudes 
${\cal M}_N^A$ and ${\cal M}_{N \rightarrow \Delta}^A$,
the factors are approximated by
the pion wave function as long as the $q\bar q$ pair has the same
spin structure of the pion. 
For spin-independent GPDs in the vector amplitudes ${\cal M}_N^V$ and
${\cal M}_{N \rightarrow \Delta}^V$, the factors are related to
the $\rho$ meson. 
In this way, the $N+N \rightarrow N+\pi+B$ reaction can be
calculated by the $N \rightarrow B$ transition part and
the $MN \rightarrow \pi N$ scattering process:
\begin{align}
{\cal M}_{NN\ra N\pi B} 
& =  {\mathscr M}_{N \rightarrow h(q\bar{q}) B} 
  \cdot {\mathscr M}_{h(q\bar{q}) N\ra \pi N} ,
\nonumber \\
& \simeq {\cal M}_{N\ra B} \cdot {\cal M}_{M N\ra \pi N}
/\phi_M (z),
\label{eqn:fact1}
\end{align}
at small momentum transfer ($|t| \ll m_N^2$). 

It is worth noting here that the discussed process is 
directly sensitive to the dependence of the GPDs on $\xi$.
Dependence on $x$ is integrated over. This is similar 
to the situation in most of the other exclusive hard
processes, for example the reaction 
$\gamma^* +p \to \pi^+ +n$.

Now, the remaining part is the description
of the $M + N \rightarrow \pi + N$ scattering.

\subsection{Parametrization of cross section for meson-nucleon
            elastic scattering}
\label{piN-parametrization}

\begin{table*}[t]
\caption{Determined parameters in Eq. (\ref{eqn:fit}).}
\label{parameter}
\centering
\begin{tabular}{@{\hspace{0.3cm}}c@{\hspace{0.7cm}}
c@{\hspace{0.7cm}}c@{\hspace{0.7cm}}c@{\hspace{0.3cm}}c}
\hline
\hline
Process
   & \multicolumn{3}{c@{\hspace{0.2cm}}}{Parameters} 
   & $\chi^2$/d.o.f.                                   \\
            & $a$ ($\rm{barn \cdot (GeV)^{2n-6}}$) 
            & $c$ ($\rm{barn \cdot (GeV)^{2n-10}}$)
            & $n-2$ (Ref.\,\cite{White94}) 
            &                                          \\
\hline
$\pi^+ p\ra\pi^+ p$  & $2.08 \pm 0.10$ 
                     & \ $1.11 \pm 0.30$ 
                     &   $6.7  \pm 0.2 $ 
                     & 0.96               \\
$\pi^- p\ra\pi^- p$  & $ 8.94 \pm 0.42$ 
                     & $10.46 \pm 1.55$ 
                     & $ 7.5  \pm 0.3$ 
                     & 0.36               \\
$\pi^+ p\ra\rho^+ p$ & \ $189 \pm 14$ \ \, 
                     & $60.0$ \ \ \ \ \ \ \ \ \ \,
                     & $8.3 \pm 0.5$ 
                     & 0.21               \\
$\pi^- p\ra\rho^- p$ & \ $218 \pm 12$ \ \, 
                     & $200.0$ \ \ \ \ \ \ \ \ \ \ \ \ 
                     & $ 8.7  \pm 1.0$ 
                     & 0.80               \\
\hline
\hline
\end{tabular}
\end{table*}

In order to calculate the cross section of meson-nucleon
elastic scattering $M p \ra \pi p$, where $p$ indicates the proton,  
we try to obtain useful parametrizations of the cross sections
by using experimental measurements.
The cross section is first expressed in terms of parameters
which are then determined by fitting $\pi p \rightarrow \pi p$ 
and $\pi p \rightarrow \rho p$ scattering data.
The momentum-transfer dependence of the cross section is typically
parametrized as \cite{White94} 
\begin{equation}
\frac{d\sigma_{M p \rightarrow \pi p} (s', t')}{dt'}
\propto a + c \, (t'-t'_0)^2 , 
\label{t90}
\end{equation}
where $a$ and $c$ are parameters, and $t'_0$ is defined by
the $t'$ value at the scattering angle $\theta_{\rm c.m.}=90^{\circ}$
in the center-of-mass (c.m.) system.
The Mandelstam variables $s'$ and  $t'$ are  given in
Eq. (\ref{eqn:mandelstam1}).
Moreover, this cross section scales with the c.m. energy squared ($s'$) as
\begin{align}
\frac{d\sigma_{M p \rightarrow \pi p} (s', t')}{dt'} 
\sim & \frac{1}{s^{'n-2}} f_{M p \rightarrow \pi p}(t'/s') 
\nonumber \\
& {\rm for}\ s'\ra\infty \  \text{ at fixed} \ \frac{t'}{s'} ,
\end{align}
according to the counting rule \cite{bf7375, chh97}.
Here, $f_{M p \rightarrow \pi p}(t'/s')$ is a function of $t'/s'$.
The factor $n$ is the total number of all interacting elementary fields,
and it is basically the sum of valence-quark numbers in the initial
and final states. 
In the meson-nucleon elastic scattering, $n=10$ and so
the energy dependence is $1/s'^{8}$. 

\begin{figure}[t]
\begin{center}
\includegraphics[width=0.30\textwidth]{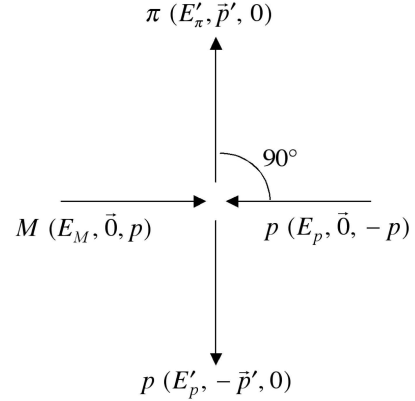}
\caption{$M p \rightarrow \pi p$ elastic scattering
          at $\theta_{c.m.} =90^{\circ}$}
\label{fig:pip90}
\end{center}
\end{figure}

The factor  $t'_0$ is defined in the c.m. frame,
in which our momentum assignment $(E,\vec{p}_{T},p_{L})$ is shown
in Fig. \ref{fig:pip90}. Here, $p_T$ and $p_L$ indicate
transverse and longitudinal momenta, respectively.
From the energy-momentum conservation,
the energies and momenta are given for the mesons and protons as
\begin{align}
& E_{M}=\frac{s'-(m_p^2 -m_{M}^2)}{2\sqrt{s'}} , \ \ 
  E_{\pi}'=\frac{s'-(m_p^2 -m_{\pi}^2)}{2\sqrt{s'}} ,
\\
& E_{p} =\frac{s'+(m_p^2 -m_{M}^2)}{2\sqrt{s'}} , \ \ 
  E_{p}'=\frac{s'+(m_p^2 -m_{\pi}^2)}{2\sqrt{s'}} , 
\\
& p  = \frac{\sqrt{s'^2 -2s'(m_p^2 +m_{M}^2) 
                   +(m_p^2 -m_{M}^2)^2}}{2\sqrt{s'}} ,
\\
& p' = \frac{\sqrt{s'^2 -2s'(m_p^2 +m_{\pi}^2) 
                   +(m_p^2 -m_{\pi}^2)^2}}{2\sqrt{s'}} . 
\end{align}
Therefore, $t'_0$ becomes
\begin{align} 
\! \! \! \! 
t'_0 & = (p_{b}-p_{d}')^2 \left. \right |_{\theta =\frac{\pi}{2}}
\nonumber \\
& = -\frac{s'^2 - s'(2 m_p^2 +m_{M}^2 +m_{\pi}^2) 
           +(m_p^2 -m_{M}^2)(m_p^2 -m_{\pi}^2)}{2s'} . 
\end{align}
In this way, the differential cross section is parametrized
in terms of the variables $s'$ and $t'$ as
\begin{equation}
\frac{d\sigma_{M p \rightarrow \pi p} (s', t')}{dt'} =\frac{1}{s'^{n-2}}
\left[ a+c(t'-t'_0)^2 \right] . 
\label{eqn:fit}
\end{equation}

\begin{figure}[t]
\begin{center}
\includegraphics[width=0.40\textwidth]{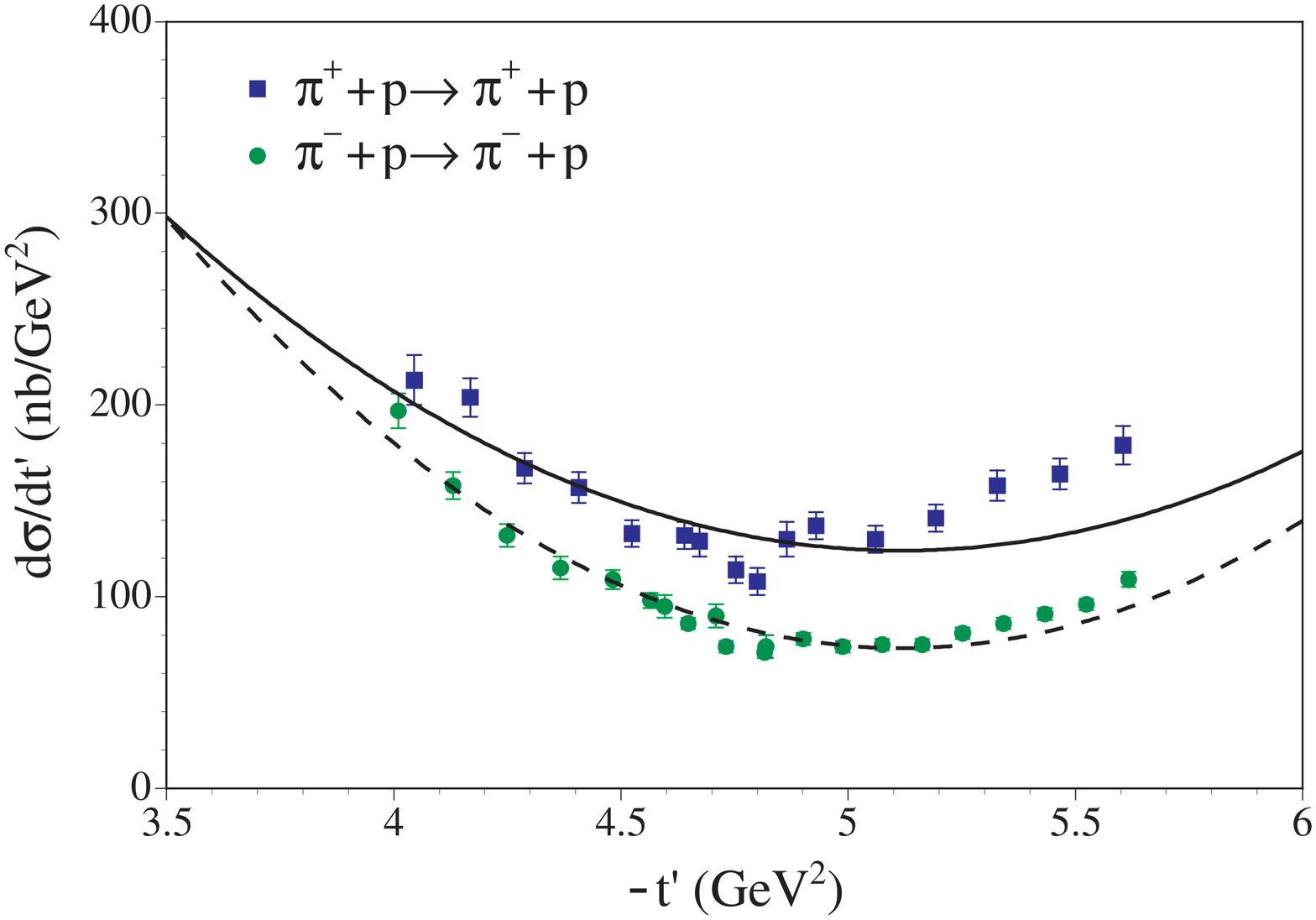}
\caption{Data of $\pi^+p \rightarrow \pi^+p$ and
             $\pi^-p \rightarrow \pi^-p$ cross sections from
         the BNL-E838 experiment \cite{White94}
         and our parametrization curves of Eq. (\ref{eqn:fit})
         with the optimized parameters.
         The solid and dashed curves are for 
         the $\pi^+p \rightarrow \pi^+p$ and
             $\pi^-p \rightarrow \pi^-p$ cross sections, 
         respectively.
         }
\label{fig:pip-data-fit}
\end{center}
\end{figure}

\begin{figure}[t]
\begin{center}
\includegraphics[width=0.40\textwidth]{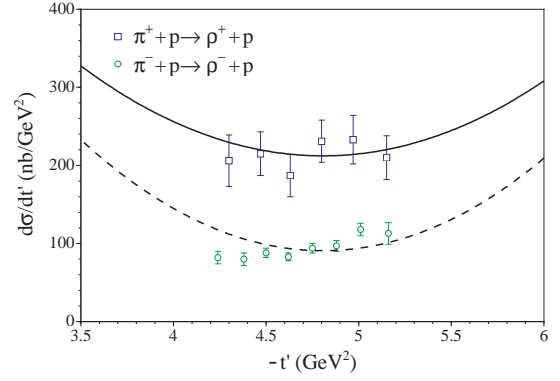}
\caption{Data of $\pi^+p \rightarrow \rho^+p$ and
             $\pi^-p \rightarrow \rho^-p$ cross sections from
         the BNL-E838 experiment \cite{White94}
         and our parametrization curves of Eq. (\ref{eqn:fit})
         with the optimized parameters.
         The solid and dashed curves are for 
         the $\pi^+p \rightarrow \rho^+p$ and
             $\pi^-p \rightarrow \rho^-p$ cross sections, 
         respectively.
         }
\label{fig:rhop-data-fit}
\end{center}
\end{figure}

Experimental data are taken from
the Alternating Gradient Synchrotron (AGS) experiment E838 \cite{White94}
with the incident pion-beam 5.9 GeV
at the Brookhaven National Laboratory (BNL).
They observed large-angle scattering cross sections for 
$\pi^+ p \rightarrow \pi^+ p$, 
$\pi^- p \rightarrow \pi^- p$, 
$\pi^+ p \rightarrow \rho^+ p$, and
$\pi^- p \rightarrow \rho^- p$.
By analyzing the scaling behavior between E755 \cite{bnl-e755} 
and E838 data, the parameter $n-2$ was determined as 6.7, 7.5, 8.3, and 8.7
in Ref. \cite{White94} as listed in Table \ref{parameter}.

We use Eq. (\ref{eqn:fit}) for describing the cross sections
by the exchange of initial and final states ($ M p \leftrightarrow \pi p$).
The parameters $a$ and $c$ are determined by fitting
the E838 data. As shown in Figs. \ref{fig:pip-data-fit}
and \ref{fig:rhop-data-fit}, there are 16, 21, 6, 8 data points for 
$\pi^+ p \rightarrow \pi^+ p$, 
$\pi^- p \rightarrow \pi^- p$, 
$\pi^+ p \rightarrow \rho^+ p$, and
$\pi^- p \rightarrow \rho^- p$, respectively. 
In Table \ref{parameter}, determined parameters are shown
and $\chi^2$ values per degrees of freedom are
$\chi^2$/d.o.f.=0.96, 0.36, 0.21, and 0.80. 
The two-parameter fit for the $\pi^\pm p \rightarrow \rho^\pm p$ was 
difficult due to a small number of data. 
Therefore, the parameter $c$ is
fixed at 60 and 200 for $\pi^+ p \rightarrow \rho^+ p$ and 
$\pi^- p \rightarrow \rho^- p$, respectively,
so as to have similar $t$-dependency to the 
$\pi^\pm p \rightarrow \pi^\pm p$ cross sections in 
Fig. \ref{fig:pip-data-fit}.
The fits are successful in reproducing the data
except for some deviations at $t \sim - 5.5$ GeV$^2$ in
Fig. \ref{fig:pip-data-fit} and
at $t \sim - 4.3$ and $-5.0$ GeV$^2$ 
for $\pi^- p \rightarrow \rho^- p$ in
$\pi^- p \rightarrow \rho^- p$ of
Fig. \ref{fig:rhop-data-fit}.
 
\section{Numerical results for cross sections}
\label{results}

\subsection{Expression for $\mathbf{N+N \rightarrow N+\pi+B}$ cross section}
\label{results-cross}

In this section we calculate the cross sections for
the $N+N \rightarrow N+\pi+B$ using the parametrizations 
for the $M+N \rightarrow \pi+N$ cross sections, and a model
for the nucleonic and $N \rightarrow \Delta$ transition GPDs.
Here, the expression for the cross section is summarized.

The $M N \ra \pi N$ cross section is given by
\begin{align}
& \! \! 
d\sigma_{M N \ra \pi N} 
= \frac{1}{4\sqrt{(p_b \cdot p_{M})^{2}
-m_N^{2} p_{M}^2}}\frac{d^{3}p_c}{2E_c (2\pi)^{3}}
\frac{d^{3}p_d}{2E_d (2\pi)^{3}}
\nonumber \\
& \! \! \! \!
\times \overline{\sum_{\lambda_b}} \sum_{\lambda_d}
    |{\cal M}_{M N \ra \pi N}|^2
    (2\pi)^4 \delta^4 (p_{b}+p_{M}-p_{c}-p_{d}) ,
\end{align}
where $\overline \sum_{\lambda_b}$ indicate the average over
the initial nucleon ($b$) spin.
The general expression for the cross section is given in
Eq. (\ref{eqn:Xsection}) for the process $N+N \rightarrow N+\pi+B$.
The factor $S$ is $S=1$ in our numerical results for 
$B=\Delta^0$, $\Delta^{++}$, or neutron, which is not identical
to the proton. By assuming the factorization and mesonic $q\bar{q}$
dominance, the matrix-element part is separated into
two processes as shown in Eq. (\ref{eqn:fact1}).
Substituting Eq. (\ref{eqn:fact1}) into Eq. (\ref{eqn:Xsection}),
we write the $N+N \rightarrow N+\pi+B$ cross section
by the $N \rightarrow B$ amplitude, namely by
the $N \rightarrow B$ GPDs, and the $M N \rightarrow \pi N$ scattering
cross section:
\begin{align}
& d\sigma _{NN \rightarrow N \pi B}
 =  \frac{\sqrt{(p_b \cdot p_{M})^{2}-m_N^{2}p_{M}^2}}
    {\sqrt{(p_a \cdot p_b)^{2}-m_N^4}}\frac{d^{3}p_e}{2E_e (2\pi)^{3}} 
\nonumber \\
& \ \ \ 
\times 
 \overline{\sum_{\lambda_a}} \sum_{\lambda_e}
 \frac{1}{[\phi_M (z)]^2} 
 |{\cal M}_{N \ra B}|^{2} 
 \, d\sigma_{M N \ra \pi N}(s',t') .
\label{eqn:X-NN}
\end{align}

We take the $z$ axis as the momentum direction of the initial 
nucleon $b$, so that our momentum assignment for $p_a$, $p_b$, and
$p_e$ is
\begin{align}
& \! \! \! \! 
p_a = (m_N, 0, 0, 0),
\nonumber \\
& \! \! \! \! 
p_b = (E_N, 0, 0, p_N),
\nonumber \\
& \! \! \! \! 
p_e = (E_B, p_B \sin \theta_e \cos \phi_e, 
              p_B \sin \theta_e \sin \phi_e, p_B \cos \theta_e),
\label{eqn:p-abe}
\end{align}
where $\theta_{e}$ and $\phi_e$ are the polar and azimuthal angles 
for the baryon $B$. 
The integral over the $B$ momentum $\vec p_e$ is written as
\begin{equation}
d^{3}p_{e} = p_e^2 dp_e d\Omega_e = p_e^2 dp_e d(\cos\theta_{e}) d\phi_{e} ,
\end{equation}

The differential cross section for $NN \rightarrow N \pi B$ 
is expressed with respect to the kinematical variables
$t$, $t'$, $y$, $\phi_{d}$, and $\phi_e$,
where the variable $y$ is defined by
\begin{equation}
y \equiv \frac{s'}{s}
 =\frac{t+m_N^2+2(m_N E_N - E_B E_N + p_B p_N \cos \theta_e)}{s} .
 \label{eqn:def-y}
\end{equation}
The variable $\phi_{d}$ is the azimuthal angle for the final nucleon $(d)$.                      
Then, integrating over $y$, $\phi_{d}$, and $\phi_{e}$, 
we finally obtain the following differential cross section: 
\begin{align}
& \frac{ d\sigma _{NN \rightarrow N \pi B} }
       {dt \, dt'}
= \int_{y_{min}}^{y_{max}} dy \ 
\frac{s}{16 \, (2 \pi)^2 \, m_N \, p_N}
\nonumber \\
& \! \! \! 
\times 
\sqrt{ \frac{(ys-t-m_N^2)^2 -4m_N^2 t }
            {(s-2m_N^2)^2   -4m_N^4} }
\, \frac{d\sigma_{M N \ra \pi N}(s'=ys,t')}{dt'}
\nonumber \\
& \! \! \! 
\times 
 \sum_{\lambda_a, \, \lambda_e}
 \frac{1}{[\phi_M (z)]^2}
 |{\cal M}_{N \ra B}|^{2} ,
\label{eqn:X-NN3}
\end{align}
where $y_{max}$ and $y_{min}$ are upper and lower bounds of the integral.
From Eq. (\ref{eqn:def-y}), $y_{max}$ is obtained by taking
$\cos \theta_e=1$. The lower bound is estimated by applying a hard
kinematical condition for $t'$ and $u'$: $|t'|, \ |u'| \ge Q_0^2$,
where the hard scale is taken in the region $Q_0^2$=2$\sim$3 GeV$^2$.
The relation $s'+t'+u'=t+2m_N^2+m_\pi^2$ leads to
\begin{equation}
y_{min} = \frac{Q_0^2 + 2 m_N^2 -t'}{s}, \ \ 
-t' \ge Q_0^2 ,
\label{eqn:ymin}
\end{equation}
where $|t|, \ m_\pi^2 \ll m_N^2$ is assumed.

We specify all used kinematical variables and constants.
The initial proton energy is given by
\begin{equation}
E_N=m_N+T_N ,
\end{equation}
where the kinetic energy of the proton beam ($T_N$) is,
for example, 30 GeV at the initial stage of J-PARC and is expected
to be 50 GeV at the later stage. Then, the c.m. energy squared is
calculated as
\begin{align}
s & = 2m_N^2 +2m_N E_N , 
\end{align}
The momentum transfer squared $t$ is defined in Eq. (\ref{eqn:mandelstam1})
and we present the cross section as a function of this variable
in Eq. (\ref{eqn:X-NN3}). 
The value of $t$ determines the energy and momentum
of the final baryon $B$: 
\begin{align}
E_B = \frac{m_N^2 + m_B^2 -t}{2 m_N}, \ \ 
p_B = \sqrt{E_B^2 - m_B^2} ,
\label{eqn:ep-b}
\end{align}
in the rest frame of the nucleon $(a)$.
The polar angle of $\vec{p}_B$ is given by
\begin{align}
\cos \theta_e & = \frac{ys -t -m_N^2 +2 (E_B-m_N) E_N}{2 \, p_B \, p_N} .
\label{eqn:theta}
\end{align}

In calculating the skewdness parameter $\xi$, we should be careful
about the reaction kinematics. We consider a reaction by
taking $z$ axis along the incident proton beam, and the GPDs
of a target hadron at rest are studied.
In the c.m. frame, the target momentum along the $z$ axis is 
negative. Therefore, the appropriate choice of
the vector $n^\mu$ should be $n_+ ^\mu$ instead of the usual
$n_- ^\mu$ in many GPD articles:
\begin{equation}
n^\mu \equiv n_+ ^{\mu} = 
  \left ( \frac{1}{\sqrt{2} P^-}, \ 0, \ 0, 
          \  \frac{1}{\sqrt{2} P^-} \right ) .
\end{equation}
The skewdness parameter $\xi$ becomes
\begin{equation}
\xi = - \frac{1}{2} n \cdot \Delta
    =    \frac{m_N -E_B + p_B \cos \theta_e}    
              {m_N +E_B - p_B \cos \theta_e} ,
\label{eqn:xi-2}
\end{equation}
by using $\Delta = p_e - p_a$ and Eq. (\ref{eqn:p-abe}).
Substituting Eq. (\ref{eqn:theta}) into this expression,
we obtain
\begin{align}
\xi = - 
\frac{ys-t-m_N^2+2(E_B-m_N)(E_N-p_N)}
     {ys-t-m_N^2+2(E_B-m_N)E_N- 2 (E_B+m_N) p_N} .
\label{eqn:xi-3}
\end{align}
The kinematical minimum $\xi_{min}$ is calculated by substituting
$y_{min}$ into the above equation, whereas the maximum $\xi_{max}$
is obtained by Eq. (\ref{eqn:xi-2}) with $\cos \theta_e=1$.

Here, it should be noted that the energy and momentum of $B$
depend on the variable $t$ as explicitly shown in Eq. (\ref{eqn:ep-b}).
The relation between $\xi$ and $y$ is obtained from
Eqs. (\ref{eqn:def-y}) and (\ref{eqn:xi-2}):
\begin{align}
y & = \frac{1}{s}  \bigg[ t + m_N^2 - 2 ( E_N -p_N ) \, E_B
\nonumber \\
  & \ \ \ \ \ \ \  
     + 2 m_N E_N - 2 \frac{1-\xi}{1+\xi} m_N \, p_N  \bigg ] ,
\label{eqn:y-xi}
\end{align}
which will be used in changing the variable $y$ for $\xi$
when presenting the numerical results for
the cross section in Sec. \ref{numerical}.

The elastic scattering factor $d\sigma_{M N \ra \pi N}(s',t') / dt'$
is calculated by using the result in Sec. \ref{piN-parametrization},
in particular Eq. (\ref{eqn:fit}) together with the parameter values
in Table \ref{parameter}.
The last ingredient is the matrix elements $M_{N \rightarrow B}$ which
are expressed by the GPDs. To be specific, we use
Eqs. (\ref{eqn:n-emplitude2}), (\ref{eqn:MV-N}), and (\ref{eqn:MA-N})
for calculating 
$\sum_{\lambda_{N}, \lambda_{N'}} |{\cal M}_{N}|^2$.
The $N \rightarrow \Delta$ matrix element
$\sum_{\lambda_{N}, \lambda_{\Delta}} |{\cal M}_{N \rightarrow \Delta}|^2$
is calculated by using
Eqs. (\ref{eqn:n-delta-emplitude2}), (\ref{eqn:mv-n-d}), 
(\ref{CFofUNPOL}), (\ref{MEofPOL}), (\ref{CFofPOL1}), (\ref{CFofPOL2}),
and (\ref{CFofPOL3}).
As we explain in the following subsection, in our numerical analysis
we estimate the meson-pole contributions to the cross sections. 
The meson-pole contributions to the GPDs contain the meson wave function
$\phi_M (z)$, which is divided out from the cross-section expression
in Eq. (\ref{eqn:X-NN3}). Therefore, a specific functional form of
$\phi_M (z)$ is not necessary for our current numerical estimates.

\subsection{Contributions from the meson poles}
\label{poles}

In this article we want to provide {\it first estimate of the order
of magnitude for the $N+N \rightarrow N+\pi+B$ cross sections in order to
establish feasibility of investigating the GPDs at hadron experimental
facilities}. For this purpose, we employ a very simple description
by meson poles. Our studies should be intended merely to provide
a starting point for future theoretical and experimental studies. 

In the ERBL region, major contributions to the GPDs come from meson poles,
particularly the pion and $\rho$-meson poles, at small $|t|$.
For numerical evaluations of the $N+N \rightarrow B +\pi +N$ cross sections,
we show explicit expressions for these pole contributions to the nucleonic
and $N \rightarrow \Delta$ transition GPDs.

First, the pion is a pseudo-scalar particle, so that it is related
to the axial part of the nucleonic matrix element. The pion pole
contributes only to the nucleonic GPD $\widetilde E (x,\xi,t)$ as 
\cite{gpv01, ppg00}
\begin{align}
\widetilde E (x,\xi,t)
= \frac{4 \, g_A \, m_N^2}{1-t/m_\pi^2} 
  \frac{\sqrt{3}}{m_\pi^2 \, f_\pi \, \xi} 
     \, \phi_\pi (z)
     \, \theta(\xi -|x|) ,
\end{align}
where $g_A$ is the axial charge of the nucleon and its value is 
$g_A = 1.2695$ \cite{pdg08}.
Here, $\widetilde E$ indicates the isovector combination
$\tilde E^u-\tilde E^d$.
In the same way, $\rho$ contributions can be estimated.
Since the $\rho$ is a vector particle, it contributes to
the vector part of the matrix element. 
The $\rho$-pole terms are given by extending 
the expression in Ref. \cite{feynman}:
\begin{align}
H (x,\xi,t)
= {1\over 1-t/m_{\rho}^2} 
    \, \frac{C_1}{g_\rho}
        \frac{\sqrt{3}}{\sqrt{2} \, f_\rho \, \xi} 
     \, \phi_\rho (z)
     \, \theta(\xi -|x|) ,
\nonumber \\
E (x,\xi,t)
= {1\over 1-t/m_{\rho}^2} 
    \, \frac{C_2}{g_\rho}
        \frac{\sqrt{3}}{\sqrt{2} \, f_\rho \, \xi} 
     \, \phi_\rho (z)
     \, \theta(\xi -|x|) ,
\end{align}
where the $C_1$ and $C_2$ are two-types of $\rho$ coupling
constants with the nucleon, 
and they are given for the isovector couplings, namely
for the isovector $\rho$, by 
\begin{align}
\frac{C_1}{g_\rho}  =\frac{e_p-e_n}{2}=\frac{1}{2} ,
\ \ 
\frac{C_2}{g_\rho}  =\frac{\kappa_p-\kappa_n}{2\cdot 2m_N}
=\frac{1.85295}{2m_N},
\end{align}
where $e_p$ ($e_n$) is the charge of the proton (neutron)
and $\kappa_p$ ($\kappa_n$) is the anomalous magnetic
moment of the proton (neutron) \cite{pdg08}.
There is no pion contribution to  $\widetilde H (x,\xi,t)$ because
the function $\widetilde H$ is chiral even, whereas the pion should
contribute to the chiral odd.

We may note that normalizations of the pion and $\rho$ wave functions
are given by
\begin{align}
1 & = \int_{-1}^1 dx \, \theta(\xi -|x|) \, 
        \frac{\sqrt{3}}{f_\pi \, \xi} 
     \, \phi_\pi (z)
\nonumber \\
  & = \int_{-1}^1 dx \, \theta(\xi -|x|) \, 
        \frac{\sqrt{3}}{\sqrt{2} \, f_\rho \, \xi} 
     \, \phi_\rho (z) ,
\end{align}
where the normalization factors are slightly different from
the ones in Ref. \cite{gpv01}.

As for the light-cone wave function of the meson $M$, we may take,
for example, asymptotic wave function of pion and $\rho$ mesons 
\cite{bl,rho}:
\begin{align}
\phi_\pi (z)  & = \sqrt{3} f_\pi  z (1-z) ,
\nonumber \\
\phi_\rho (z) & = \sqrt{6} f_\rho z (1-z) .
\end{align}
The normalization of these wave function is uniquely fixed
in QCD \cite{bl,rho}. For the pion wave function it is given by 
\begin{equation}
\int dz \phi_\pi (z) 
     = \int \frac{dz \, d^2 k_\perp}{16\pi^3} \psi_\pi (z,\vec k_\perp)
     = \frac{f_\pi}{2 \sqrt{3}} ,
\end{equation}
where $f_\pi$ is the pion decay constant for 
$\pi^+ \rightarrow \mu^+ \nu_\mu$.
By noting the difference of $\sqrt{2}$ from the value in 
Ref. \cite{pdg06,pdg08}, it is $f_\pi$=92.4 MeV.
The $\rho$ decay constant $f_\rho$ is determined by the decay
$\rho^0\rightarrow e^+ e^-$, and it is given by \cite{rho,pdg08}
\begin{equation}
f_\rho =\sqrt{\frac{3 m_\rho \Gamma_{\rho^0 \rightarrow e^+e^-}}
                   {8 \pi \alpha^2}}
       =110.6 \ \text{MeV}.
\end{equation}
where $m_\rho$ is the $\rho$ mass, 
$\alpha$ is the fine structure constant,
and $\Gamma_{\rho^0 \rightarrow e^+e^-}$ is the $\rho^0 \rightarrow e^+e^-$
decay width.

Next, pion and $\rho$ contributions are shown for the $N \rightarrow \Delta$
transition GPDs. The pion contributes to the axial part, particularly
the GPD $\widetilde H_2$ \cite{gpv01,h-notation}, is given in the same way by
\begin{align}
\widetilde{H}_2 (x,\xi,t)
& = \frac{3 \, g_A \, m_N^2}{(1 -t/m_{\pi}^2) \, m_{\pi}^2 \, \xi \, f_\pi}
\, \phi_\pi (z) \, \theta(\xi - |x|) .
\end{align}
The $\rho$-meson contribution to the vector part is expressed by
the magnetic $N \rightarrow \Delta$ transition as
\begin{align}
H_M (x,\xi,t)
= \frac{G^*_M}{1-t/m_{\rho}^2} 
    \,  \frac{\sqrt{3}}{\sqrt{2} \, f_\rho \, \xi} 
     \, \phi_\rho (z)
     \, \theta(\xi -|x|) ,
\end{align}
where $G^*_M$ is the transition magnetic moment \cite{js73}.
Experimental measurements indicate $G^*_M \approx 3$ \cite{ujs07},
which is close to an expectation $G^*_M = (\mu_p - \mu_n)/\sqrt{2}=3.32757$
\cite{gpv01}, where $\mu_p$ and $\mu_n$ are magnetic moments for
the proton and the neutron, respectively.
In our numerical analysis, we use the value $G_M^* =3.3$
for the transition magnetic moment.

\subsection{Numerical results}
\label{numerical}

From measurements of the $N+N \rightarrow B +\pi +N$ cross sections,
the GPDs should be extracted from the data. There is no such
data at this stage, so that we need to provide an order of magnitude
estimates of the cross sections for experimental proposals at hadron
experimental facilities, for example, J-PARC and 
GSI-FAIR \cite{j-parc, gsi-fair}.
Since large contributions should come from pion- and $\rho$-pole terms,
such pole contributions are evaluated in this work. 
However, one should note that the following numerical evaluations 
based on the meson-pole terms are just rough estimates of
the cross sections. 
In particular, there exists an extra $t$ dependence in the GPDs
at large $|t|$ (like form factors at large $|t|$) 
in addition to the pole $t$-dependence, and it is neglected in
this work. Conversely, the actual GPDs need to be extracted
from future measurements once the data are taken.
Then, they should be used for investigating the spin structure 
of the nucleon, especially orbital angular momentum effects, 
and electromagnetic properties of the $N \rightarrow \Delta$ 
transition as explained in Sec. \ref{n-gpds-prop}
and Sec. \ref{n-delta-gpds-prop}.

In the case when incident particles are unpolarized and
polarization of the final particles is averaged over, 
we predict the same cross section for analogous reactions initiated 
by antiprotons. Interference of axial and vector terms survives 
if the target proton is transversely polarized.
Interestingly, the sign of the interference effect
for incident protons and antiprotons depends on relative phase
of the $A_{\pi N\to \pi N}$ and $A_{\rho N\to \pi N}$ amplitudes. 
In leading QCD, diagrams for this process correspond to
real amplitudes, so the relative phase could be either $1$ or $-1$.
So, in principle, by combining measurements with proton
and antiproton beams, one can measure this phase.

\begin{figure}[b]
\begin{center}
\includegraphics[width=0.40\textwidth]{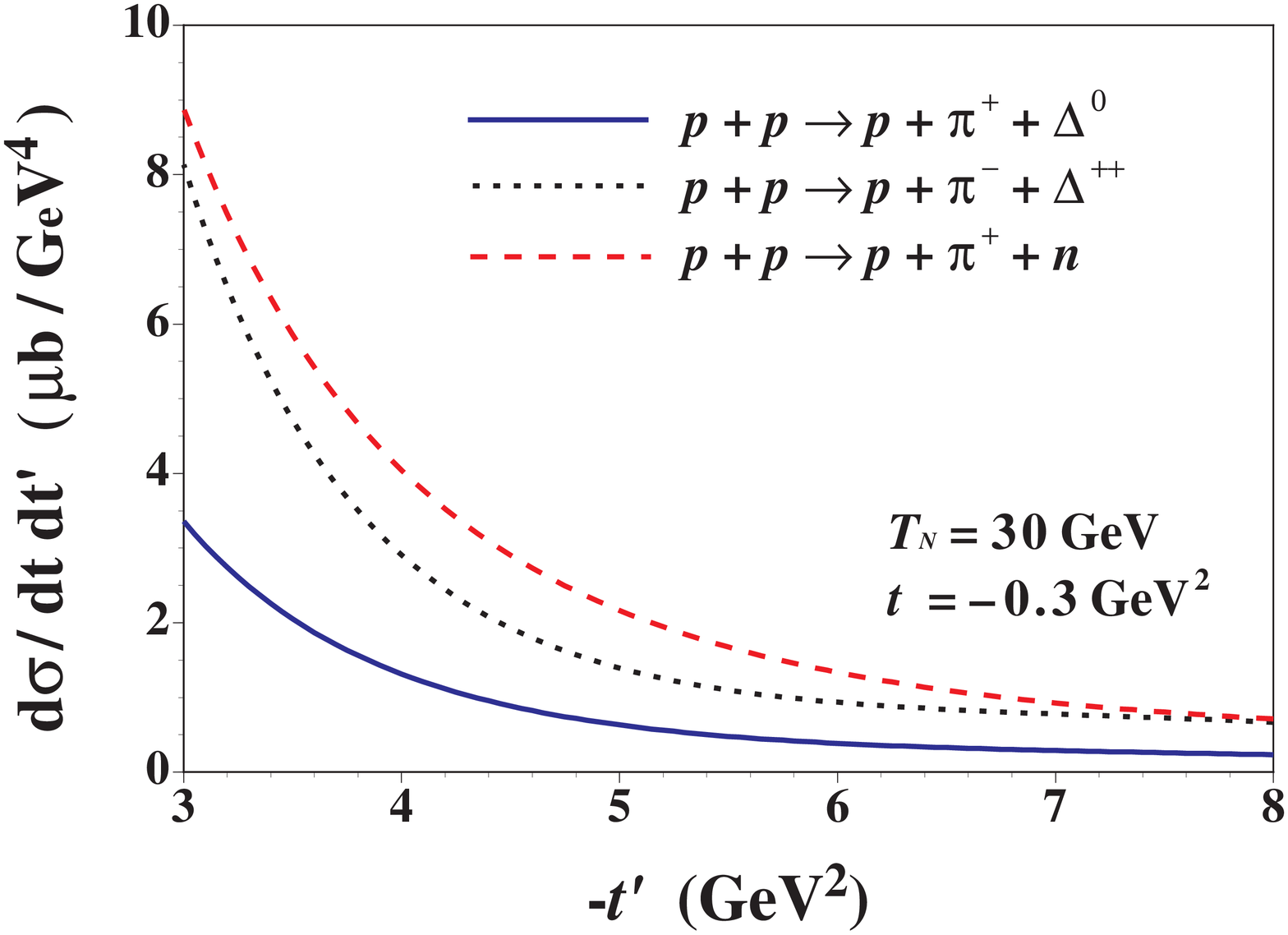} \\
\vspace{0.3cm}
\includegraphics[width=0.40\textwidth]{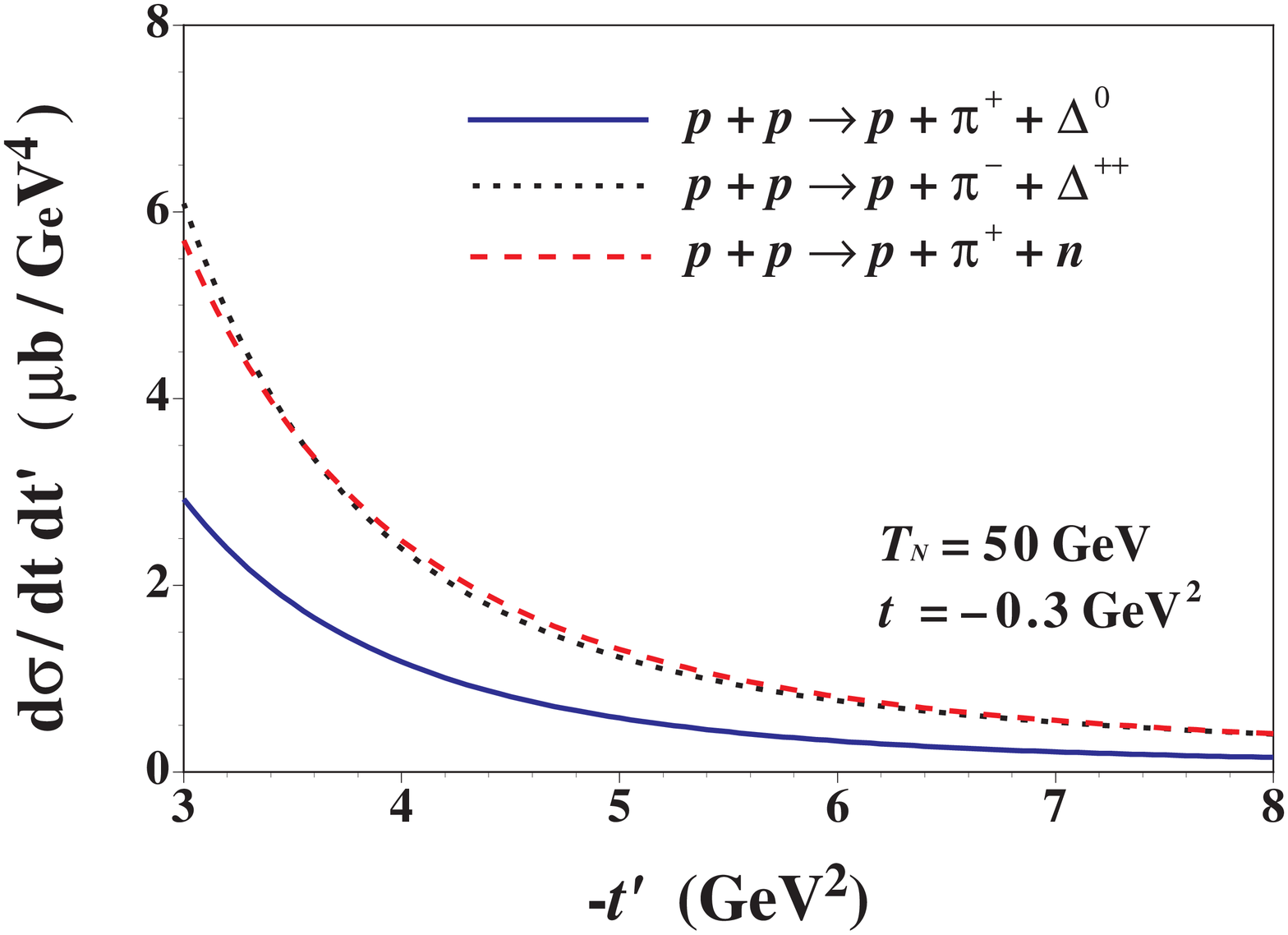}
\caption{Differential cross section as a function of $t'$. 
The incident proton-beam energy is 30 (50) GeV in the upper
(lower) figure, and the momentum transfer is $t=-0.3$ GeV$^2$.
The solid, dotted, and dashed curves indicate the cross sections
for $p + p \rightarrow p + \pi^+ + \Delta^0$,
$p + p \rightarrow p + \pi^- + \Delta^{++}$, and
$p + p \rightarrow p + \pi^+ + n$, respectively.
}
\label{fig:cross-30-50gev}
\end{center}
\end{figure}

\begin{figure}[b]
\begin{center}
\includegraphics[width=0.40\textwidth]{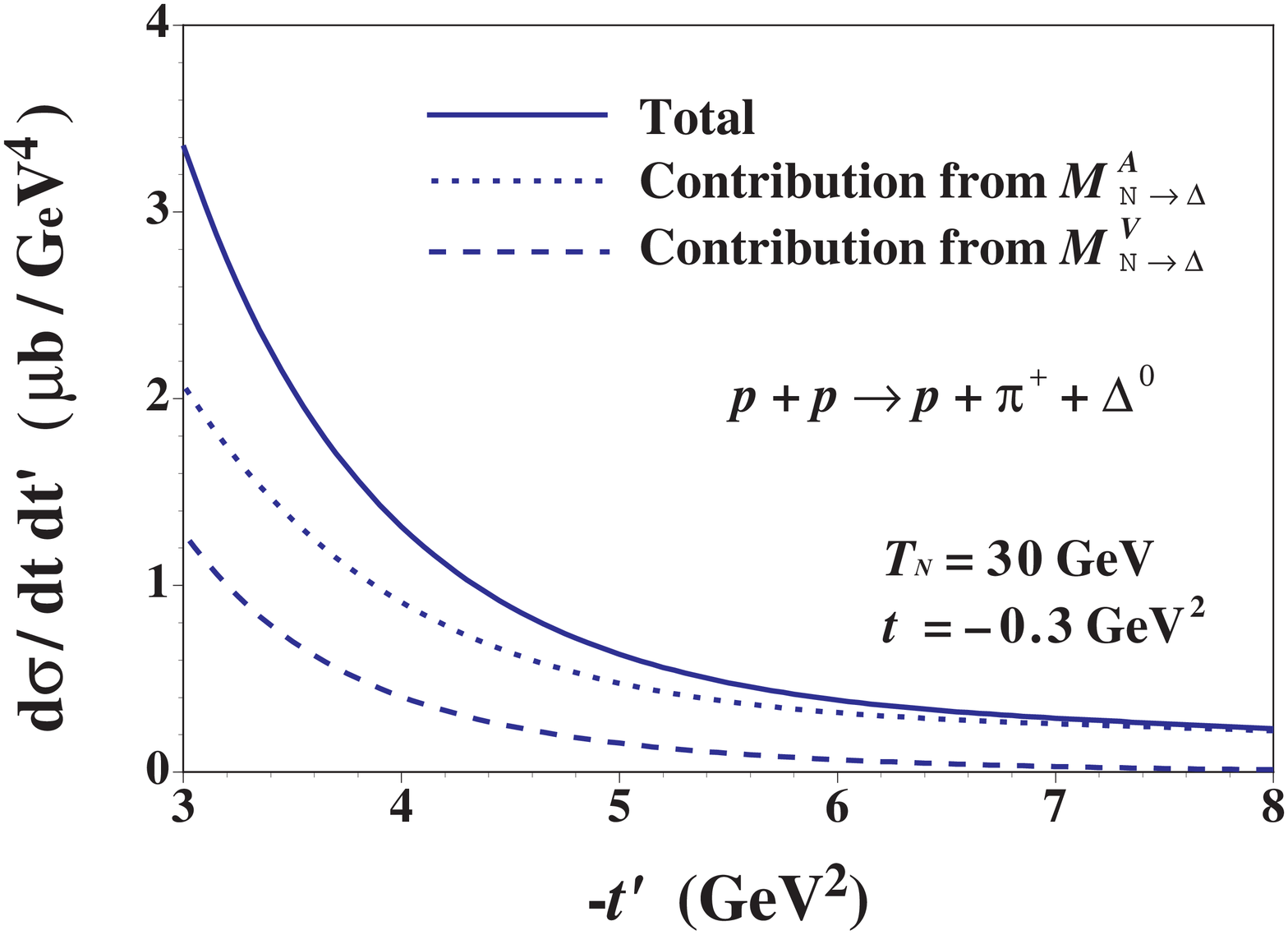} \\
\vspace{0.3cm}
\includegraphics[width=0.40\textwidth]{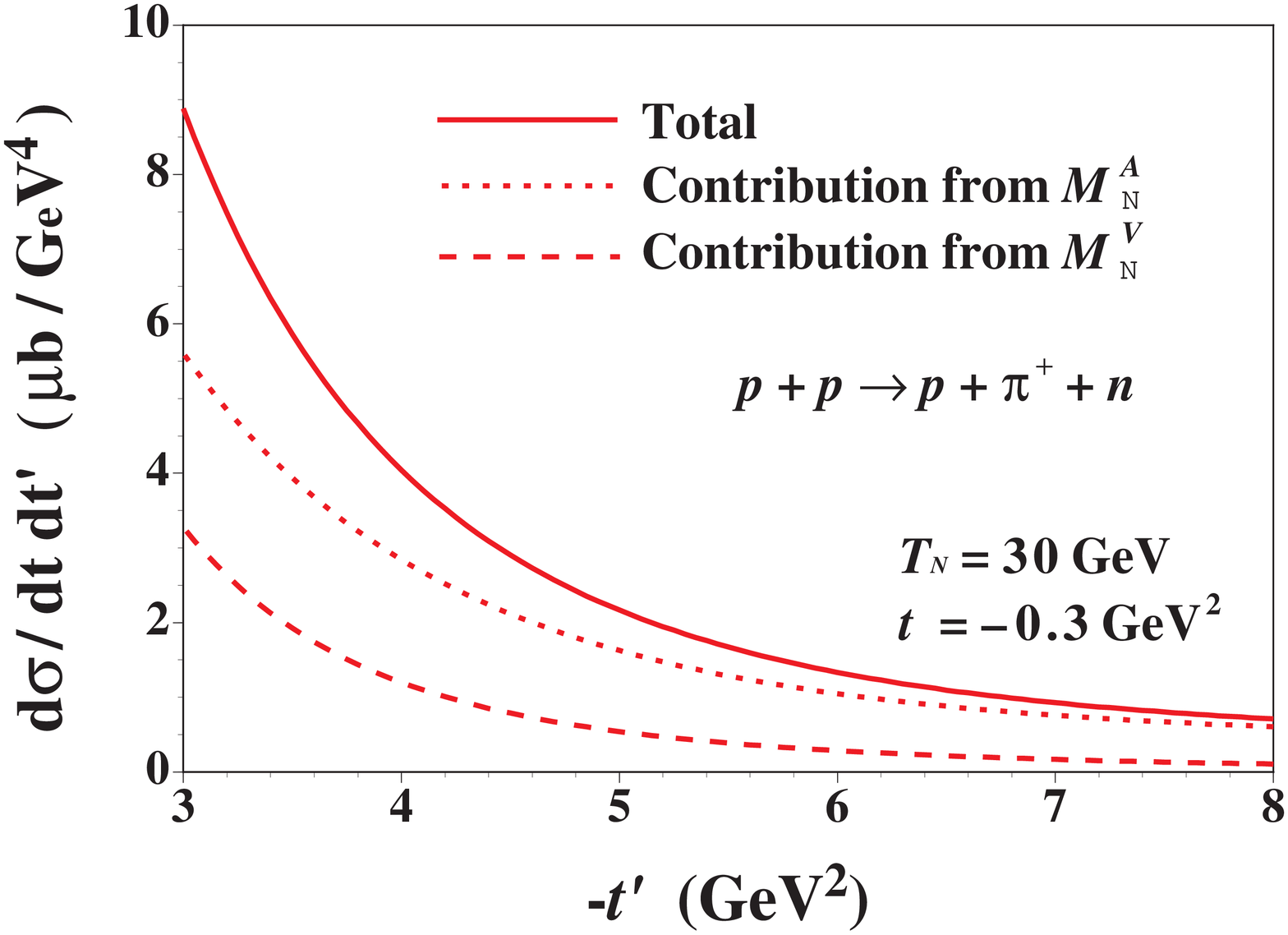}
\caption{Differential cross section as a function of $t'$. 
The incident proton-beam energy is 30 GeV, 
and the momentum transfer is $t=-0.3$ GeV$^2$.
The upper (lower) figure indicates the cross section for
the process
$p + p \rightarrow p + \pi^+ + \Delta^0$
($p + p \rightarrow p + \pi^+ + n$).
The solid, dotted, and dashed curves indicate the cross
sections for the total, axial-vector ($\pi$) contribution, 
vector ($\rho$) contribution, respectively.}
\label{fig:cross-delta-n}
\end{center}
\end{figure}

We show the numerical results for
the following three cross sections
\begin{itemize}
\vspace{-0.15cm}
\item[(1)] $p + p \rightarrow p + \pi^+ + \Delta^0$,
\vspace{-0.15cm}
\item[(2)] $p + p \rightarrow p + \pi^- + \Delta^{++}$,
\vspace{-0.15cm}
\item[(3)] $p + p \rightarrow p + \pi^+ + n$,
\vspace{-0.15cm}
\end{itemize}
in Fig. \ref{fig:cross-30-50gev}
for the kinetic energy of the proton beam at 30 and 50 GeV.
We focus on these processes for several reasons:
they are among the simplest to measure experimentally,
they are expressed through the GPDs which are best studied theoretically,
and they are  expressed through elementary large angle meson-nucleon
reactions which were studied experimentally.

Also, in these processes  the intermediate state
with the vacuum quantum number $0^+$ does not contribute
as explained in Sec. \ref{cross-section} which further
simplifies the description.
The cross sections are calculated at 
the momentum transfer $t=-0.3$ GeV$^2$, and
the momentum cutoff in Eq. (\ref{eqn:ymin}) is taken
as $Q_0^2$=3 GeV$^2$. 
The cross sections are decreasing function of $-t'$, and 
they are of the order of $\mu \,$barn/GeV$^2$.
Hence it appears  feasible to accumulate large statistics 
for these reactions over the range of $t'$ for several incident
energies and check the suggested mechanism of the reaction.
Therefore it should be in principle possible
to measure the GPDs in the $N + N \rightarrow N + \pi^+ + B$
reactions. The $\Delta$-production cross sections are as large
as the nucleonic one, which indicates that $N \rightarrow \Delta$
transition GPDs could be also studied as well as the nucleonic
GPDs. At $T_N=50$ GeV, the magnitude of the cross sections becomes
slightly smaller; however, the larger beam energy is valuable
for extending the kinematical region of $t$ (and $\xi$ as shown later)
and testing the mechanism of the reaction in particular by comparing
the cross sections for the same $s',t'$ and different $E_N$.
    
Next, the pion and $\rho$-meson contributions to
the cross sections are explicitly shown for the processes
$p + p \rightarrow p + \pi^+ + n$ and
$p + p \rightarrow p + \pi^+ + \Delta^0$
in Fig. \ref{fig:cross-delta-n} at the 30 GeV
beam energy and the momentum transfer $t=-0.3$ GeV$^2$.
The results indicate that the contribution from the $\rho$-like
vector GPD terms are smaller than the pion-like 
axial-vector terms in both process $p + p \rightarrow p + \pi^+ + n$
and $p + p \rightarrow p + \pi^+ + \Delta^0$.
Especially at large $-t'$, the $\rho$ contributions are much
smaller than the pion ones. It is expected in general because
cross sections should be suppressed for heavier-meson
intermediate states.
Therefore, it is important first to understand 
the axial-vector (quark spin independent) GPDs 
for estimating the magnitude of the cross sections.
However, both the axial-vector and vector (spin dependent)
GPDs should be calculated for estimating the cross sections.
Once experimental data are obtained, they should
be valuable for determining both spin independent and 
spin dependent GPDs for the nucleon and 
for the $N \rightarrow \Delta$ transition.

\begin{figure}[t]
\begin{center}
\includegraphics[width=0.40\textwidth]{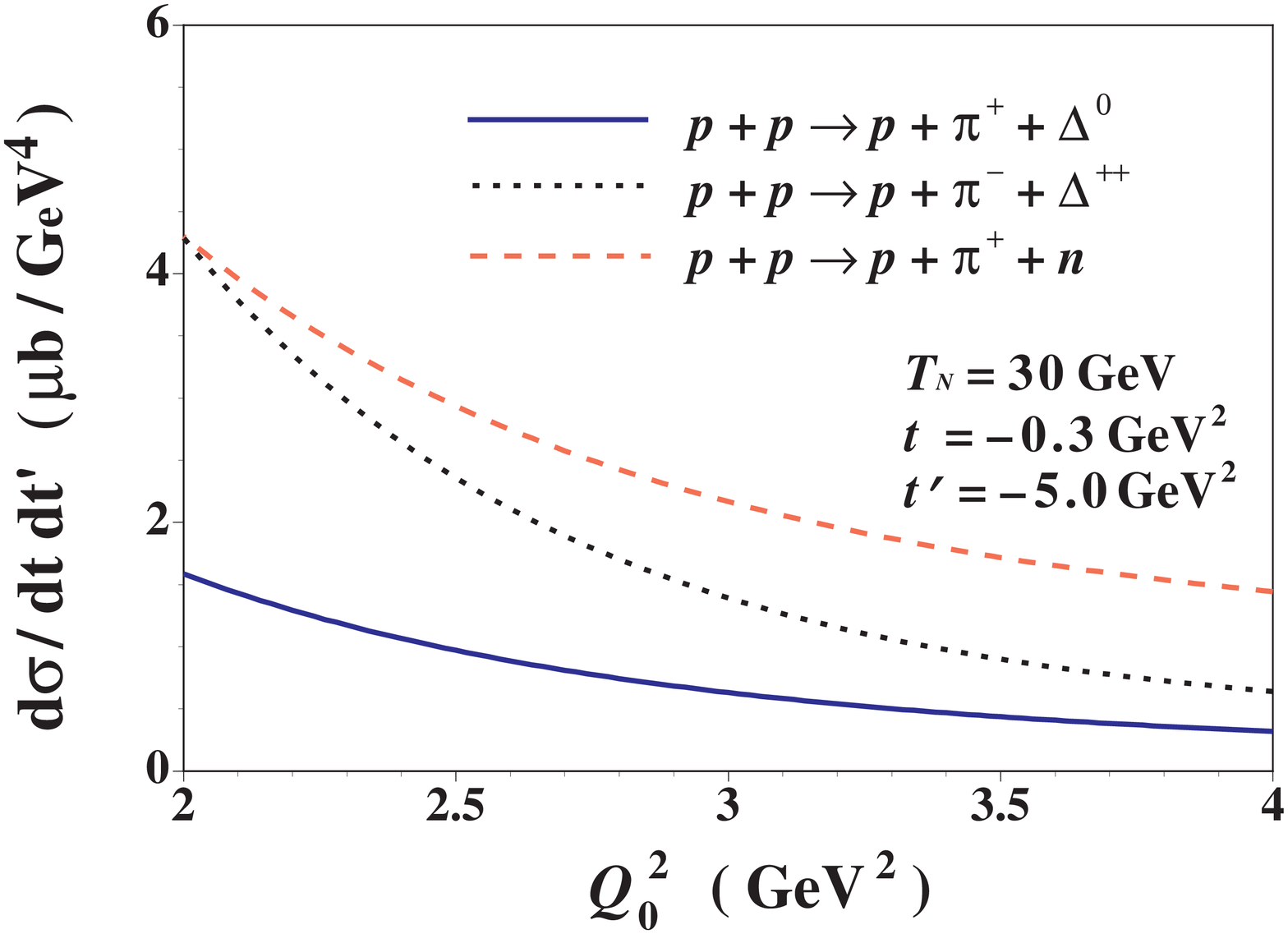}
\caption{Cutoff ($Q_0^2$) dependence of differential
cross section.
The incident proton-beam energy is 30 GeV, and 
momentum transfers are fixed at $t=-0.3$ GeV$^2$ and
$t'=-5$ GeV$^2$.
The solid, dotted, dashed curves indicate the cross sections
for $p + p \rightarrow p + \pi^+ + \Delta^0$,
$p + p \rightarrow p + \pi^- + \Delta^{++}$, and
$p + p \rightarrow p + \pi^+ + n$, respectively.
}
\label{fig:cross-cut}
\end{center}
\end{figure}

The cross-section results are presented in Figs. \ref{fig:cross-30-50gev}
and \ref{fig:cross-delta-n} for  the hard momentum cutoff 
$Q_0^2$=3 GeV$^2$ for $t'$ and $u'$  in Eq. (\ref{eqn:ymin}). 
We show the cutoff dependence of the cross sections
in Fig. \ref{fig:cross-cut} within 
the range $2$ GeV$^2$$\le Q_0^2 \le 4$ GeV$^2$.
The cross sections depend strongly
on the cutoff if it is taken low enough 
$\sim 2$ GeV$^2$. As the momentum transfers 
$|t'|$ and $|u'|$ become smaller, 
soft contributions become larger. 
It will be important to study  process as a function of $-t'$
to determine minimal value of the cutoff 
which is necessary for using discussed hadronic processes 
for determination of the nucleonic and $N\rightarrow \Delta$
transition GPDs.

\begin{figure}[t]
\begin{center}
\includegraphics[width=0.40\textwidth]{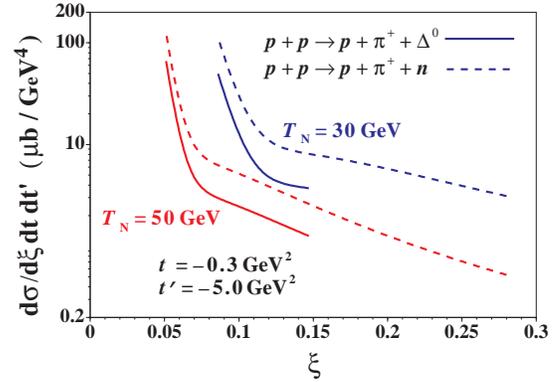}
\caption{
Differential cross sections are shown as a function of $\xi$
at the incident proton-beam energy 30 (or 50) GeV, 
$t=-0.3$ GeV$^2$, and $t'=-5$ GeV$^2$.
The solid and dashed curves indicate the cross sections
for $p + p \rightarrow p + \pi^+ + \Delta^0$ and
$p + p \rightarrow p + \pi^+ + n$, respectively.
The two upper (lower) curves indicate the cross sections
at $T_N=30$ (50) GeV.
There is a restriction on the kinematical range,
$\xi_{min}<\xi<\xi_{max}$,
which is explained in the text.
}
\label{fig:cross-xi}
\end{center}
\end{figure}

\begin{figure}[b]
\begin{center}
\includegraphics[width=0.40\textwidth]{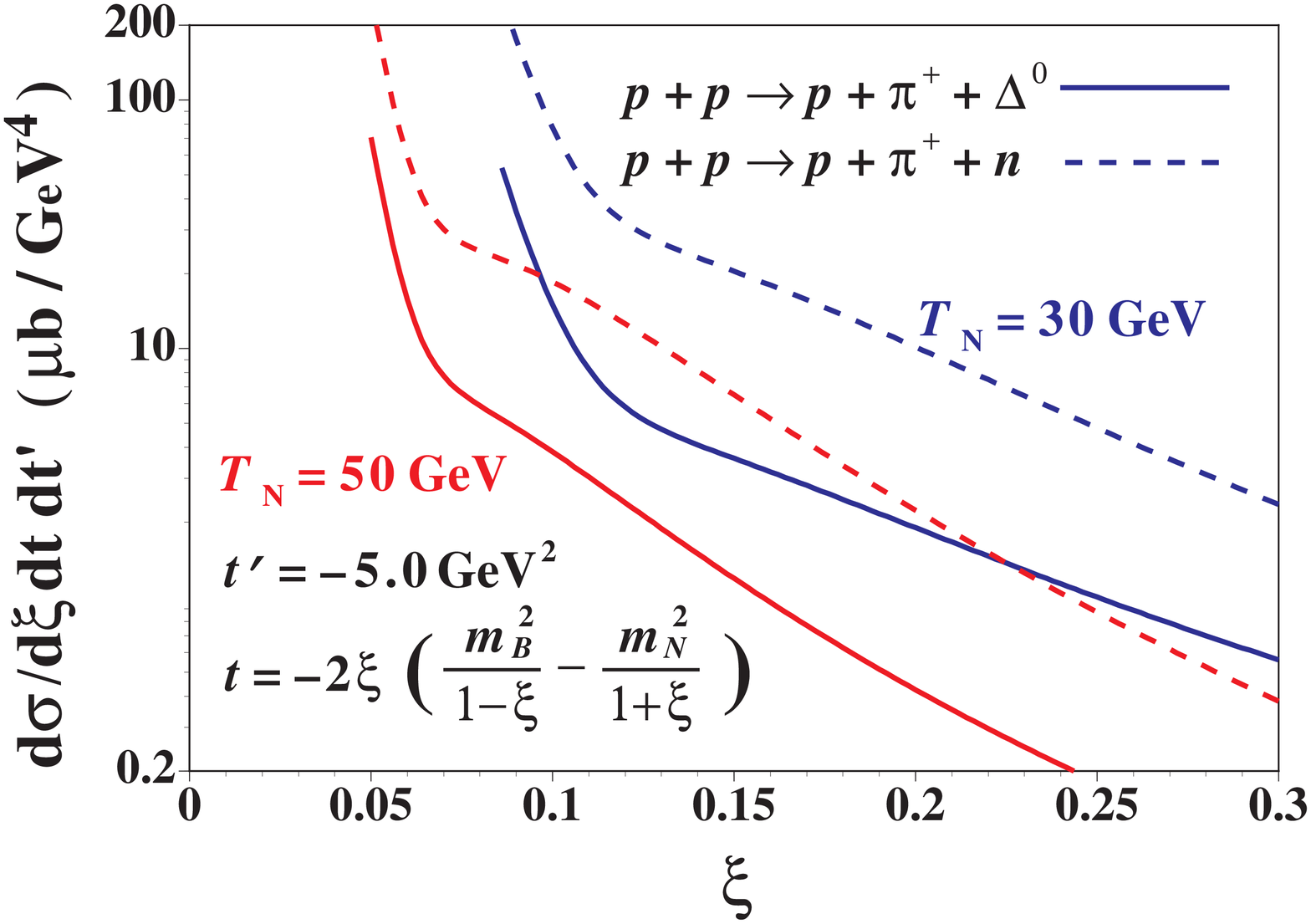}
\caption{
Differential cross sections are shown as a function of $\xi$
at the incident proton beam energy is 30 (or 50) GeV 
and $t'=-5$ GeV$^2$. The notations are the same in 
Fig. \ref{fig:cross-xi}.
The momentum transfer $t$ is taken at $\cos \theta_e=1$
as a function of $\xi$.
}
\label{fig:cross-xi-tmin}
\end{center}
\end{figure}

In connection with the studies of GPDs, it is instructive 
to present the cross sections as a function of $\xi$ 
without integrating them over the variable $y$.
Using the relation between the variables $y$ and $\xi$
in Eq. (\ref{eqn:y-xi}), we calculate the differential cross
sections $d\sigma /d\xi dt dt'$.
In Fig. \ref{fig:cross-xi}, the cross sections are shown 
as a function of $\xi$ at fixed $t$ and $t'$
($t=-0.3$ GeV$^2$ and $t'=-5$ GeV$^2$).
It should be noted that the kinematical range of $\xi$ is
limited ($\xi_{min}<\xi<\xi_{max}$) as explained below
Eq. (\ref{eqn:xi-3}). The minimum $\xi_{min}$ comes from
the hard kinematical condition for $|t'|$ and $|u'|$,
and the minimum $\xi_{min}$ is due to the minimum polar
angle for the final baryon $B$ ($\theta_e=0$).
Because of the larger masses for the $\Delta$ than the one
for the nucleon, the $\xi$ range is more restricted:
$\xi_{max}=(m_N-E_B+p_B)/(m_N+E_B-p_B)$
for the $\Delta$.
At larger beam energy ($T_N=50$ GeV), the smaller-$\xi$ region
is probed by the reactions. Therefore, 50-GeV experiments
are valuable for extending the kinematical regions 
(in $\xi$ and also $t$) of the GPD measurements 
in addition to the 30-GeV experiments.
Note in passing that the use of 50 GeV proton beam would 
allow to produce secondary beams of mesons and perform 
precision measurements of meson-nucleon large angle scattering
processes which enter in the analysis of the $2\to 3$ processes
and of interest on their own.

Next, the same cross sections are calculated at different
momentum transfer points for $t$. At a given $\xi$,
there is a restriction for the variable $t$ at $\theta_e=0$,
and it is given by 
$(-t)_{min} = 2 \xi [ m_B^2 / (1-\xi) - m_N^2 / (1+\xi) ]$.
The cross sections are calculated at this point of $t$
for each $\xi$ in Fig. \ref{fig:cross-xi-tmin}.
Then, the cross-section range is extended to larger $\xi$.
Measuring these cross sections at future facilities
at J-PARC and GSI-FAIR, we should be able to obtain
information on the GPDs in the ERBL region.

In this article, we discussed processes $N+N \rightarrow N+\pi+B$
($B$=$N$ or $\Delta$) for investigating the GPDs at the hadron
facilities. In addition to these GPD studies for the nucleon
and $\Delta$, strange-particle productions are also quite interesting,
and they can be calculated along the same line.
For example, the process $p + p \to \Lambda + p+ K^+$ is
likely to have a cross section comparable to that of
the reactions we discussed in this work
as the smaller $K\Lambda N$ vertex is to some extent
compensated by larger cross section of $K^+p\to K^+p$
scattering.

\section{Summary}
\label{summary}

We introduced a new class of hard $2\to 3$ branching hadronic processes  
and investigated a possibility that the generalized parton
distributions for the nucleon and the $N \rightarrow \Delta$
transition could be investigated in the reaction
$N+N \rightarrow N+\pi+B$ where $B$ is the nucleon or $\Delta$.
First, we argued that the process is factorized into two blocks, 
$N \rightarrow B$ and meson-nucleon ($MN$) scattering by imposing
hard scattering condition for the meson-nucleon block. 
The $N \rightarrow B$ part is expressed in terms of 
the nucleonic and the $N \rightarrow \Delta$ transition GPDs, and
the $M N$ scattering part is parametrized so as to explain
the experimental elastic scattering cross sections.
The GPDs can be studied in the ERBL region by such exclusive
reactions.

Using the meson-pole model for the GPDs, we estimated the magnitude
of the cross section of several reactions $pp\to p \pi N(\Delta)$.
We found a cross sections are  in the range measurable at
the high-energy hadron facilities.   
Experimental studies will be possible in two kinematics: 
one corresponding to the baryon $B$ been slow in the rest frame,
and another when $B$ originates from the projectile (cross sections 
for the two kinematics are equal and we presented only one in our plots). 
Experimental investigations in the first kinematics would 
require both forward and recoil detectors, while in the second
kinematics a forward multiparticle detector may be sufficient. 
Our results indicate that both the nucleonic and
the $N \rightarrow \Delta$ transition GPDs could be measured
in hadron reactions.
Future high-energy hadron facilities such as J-PARC and GSI-FAIR 
will provide opportunities to investigate various 
interesting aspects of the GPDs in a way complementary to 
the measurements using lepton scattering.
These measurements will be valuable for finding the origin of
the nucleon spin, especially on the orbital-angular-momentum
contribution. 
Note that the studied GPDs probe the isovector part
of the quark angular momenta
both in the $p\to n$ and $N \rightarrow \Delta$ channels. 


\begin{acknowledgements}
The authors thank M. V. Polyakov, S. Sawada, 
and M. Vanderhaeghen for valuable suggestions and comments. 
This research has been partially supported by 
the Research Program of Hayama Center for Advanced Studies
of Sokendai and by the US DOE Contract Number DE-FG02-93ER40771. 
MS would like to thank KEK for hospitality during the visit
when this project has started.
\end{acknowledgements}




\begin{thebibliography}{00}
\bibitem{Collins:1989gx} For a review, see
   J. C. Collins, D. E. Soper, and G. Sterman,
       Adv. Ser. Direct. High Energy Phys. {\bf 5}, 1 (1988)
       [arXiv:hep-ph/0409313].
\bibitem{ppdfs} For a recent situation, see
   S. E. Kuhn, J.-P. Chen, and E. Leader, arXiv:0812.3535 [hep-ph].
   to be published in Progress in Particle and Nuclear Physics;
   M. Hirai and S. Kumano,  Nucl. Phys. B {\bf 813}, 106 (2009).
\bibitem{deltaG}
   A. Airapetian {\it et al.} (HERMES Collaboration),
          Phys. Rev. Lett. {\bf 84}, 2584 (2000);
   B. Adeva {\it et al.} (Spin Muon Collaboration (SMC)), 
          Phys. Rev. D {\bf 70}, 012002 (2004);
   A. Adare {\it et al.} (PHENIX collaboration), 
          Phys. Rev. D {\bf 76}, 051106 (2007).     
\bibitem{DVCSfact}
   D. Muller {\it et al.}, Fortsch. Phys. {\bf 42}, 101 (1994);
   X.-D. Ji,      Phys. Rev. D {\bf 55}, 7114 (1997);
   A. V. Radyushkin, Phys. Rev. D {\bf 56}, 5524 (1997);
   J. J. Collins and A. Freund, Phys. Rev. D {\bf 59}, 074009 (1999).
\bibitem{Brodsky:1994kf}
  S. J. Brodsky, L. Frankfurt, J. F. Gunion, A. H. Mueller, and M. Strikman,
          Phys. Rev.  D {\bf 50}, 3134 (1994).
\bibitem{Collins:1997hv}
  J. C. Collins, L. Frankfurt, and M. Strikman,
          arXiv: hep-ph/9709336.
\bibitem{transverse} 
  M. Burkardt, Phys. Rev. D {\bf 62}, 071503 (2000);
                            {\bf 66}, 119903 (2002) (errattum);
  M. Diehl, Eur. Phys. J. C {\bf 25}, 223 (2002);
                            {\bf 31}, 277 (2003) (erratum);
  J. P. Ralston and B. Pire, Phys. Rev. D {\bf 66}, 111501 (2002).
\bibitem{annual}For a review, see 
   L. Frankfurt, M. Strikman, and C. Weiss,
         Ann. Rev. Nucl. Part. Sci.  {\bf 55}, 403 (2005).
\bibitem{ji97} X.-D. Ji, Phys. Rev. Lett. {\bf 78}, 610 (1997);
                         Ann. Rev. Nucl. Part. Sci. {\bf 54}, 413 (2004). 
\bibitem{gpv01}   K. Goeke, M. V. Polyakov, and M. Vanderhaeghen, 
                  Prog. Part. Nucl. Phys.  {\bf 47}, 401 (2001). 
\bibitem{diehl03} M. Diehl, Phys. Rept. {\bf 388}, 41 (2003);
                  A. V. Belitsky and A. V. Radyushkin, 
                  Phys. Rept. {\bf 418}, 1 (2005). 
\bibitem{bochum} J. Ossmann, M.V. Polyakov, P. Schweitzer, D. Urbano, 
                       and K. Goeke, Phys. Rev. D {\bf 71}, 034011 (2005). 
\bibitem{osaka} M. Wakamatsu and H. Tsujimoto, 
                     Phys. Rev. D {\bf 71}, 074001 (2005); 
                M. Wakamatsu, Phys. Rev. D {\bf 72}, 074006 (2005);
                                           {\bf 79}, 014033 (2009); 
                M. Wakamatsu and Y. Nakakoji, 
                     Phys. Rev. D {\bf 74}, 054006 (2006). 
\bibitem{chiral-p-08} M. Dorati, T. A. Gail, and T. R. Hemmert,
                         Nucl. Phys. A {\bf 798}, 96 (2008). 
\bibitem{lattice} D. Brommel {\it et al.}, (QCDSF-UKQCD Collaboration),
                         PoS LAT2007, 158 (2007);
                  P. Hagler {\it et al.} (LHPC Collaboration), 
                         Phys. Rev. D {\bf 77}, 094502 (2008). 
\bibitem{aligned-jet} A. Freund, M. McDermott, and M. Strikman,
                         Phys. Rev. D {\bf 67}, 036001 (2003).
\bibitem{gpd-paramet} S. Ahmad, H. Honkanen, S. Liuti, and S. K. Taneja,
                        Phys. Rev. D {\bf 75}, 094003 (2007);
                      O. V. Selyugin and O. V. Teryaev,
                         Phys. Rev. D {\bf 79}, 033003 (2009). 
\bibitem{gpd-exp} S. Chen {\it et al.} (CLAS Collaboration),
                         Phys. Rev. Lett. {\bf 97}, 072002 (2006);
                  F. H. Heinsius (COMPASS Collaboration), pp. 1121-1124 in
                  Proceedings of the 5th International Workshop on 
                  Deep-Inelastic Scattering and Related Subjects (2007).
\bibitem{recent-gpds} Recent studies are explained in 
                  M. Guidal, Prog. Part. Nucl. Phys. {\bf 61}, 89 (2008);
                  C. Weiss, arXiv:0902.2018 [hep-ph].
\bibitem{gpd-pp-l+l-} E. R. Berger, M. Diehl, and B. Pire,
                          Phys. Lett.  B {\bf 523}, 265 (2001).
\bibitem{fpsv9800} L. L. Frankfurt, M. V. Polyakov, and M. Strikman, 
                         hep-ph/9808449;
                 L. L. Frankfurt, M. V. Polyakov, M. Strikman, 
                 and M. Vanderhaeghen, Phys. Rev. Lett. {\bf 84}, 2589 (2000). 
\bibitem{NDelta-form} D. Drechsel and T. Walcher,
                           Rev. Mod. Phys. {\bf 80}, 731 (2008);
                      B. Krusche and S. Schadmand,
                           Prog. Part. Nucl. Phys. {\bf 51}, 399 (2003). 
\bibitem{hkmm87} L. Heller, S. Kumano, J. C. Martinez, and E. J. Moniz,
                     Phys. Rev. C {\bf 35}, 718 (1987);
                 A. Bosshard {\it et al.}, Phys. Rev. D {\bf 44}, 1962 (1991). 
\bibitem{NDelta-E2} B. Julia-Diaz, T.-S. H. Lee, T. Sato, and L. C. Smith,
                       Phys. Rev. C {\bf 75}, 015205 (2007);
   A. I. Machavariani and A. Faessler, 
              Phys. Rev. C {\bf 72}, 024002 (2005);
   A. J. Buchmann, J. A. Hester, R. F. Lebed, 
              Phys. Rev. D {\bf 66}, 056002 (2002);
   S. Kumano, Phys. Lett. B {\bf 214}, 132 (1988);
              Nucl. Phys. A {\bf 495}, 611 (1989).
\bibitem{stardust} 
   M. I. Strikman and L. L. Frankfurt, 
            pp. 211-220 in Proceedings of the 7th International Conference
            on the Structure of Baryons (1995);  
   M. Strikman and M. Zhalov, Nucl. Phys. A {\bf 670}, 135 (2000).  
\bibitem{Frankfurt:2002kz}
   L. Frankfurt, M. V. Polyakov, M. Strikman, D. Zhalov, and M. Zhalov,
           arXiv:hep-ph/0211263,
           pp. 361-368 in Proceedings of of Exclusive Processes 
           at High Momentum Transfer, Newport News, Virginia, 15-18 May 2002.
\bibitem{erbl} A. V. Efremov and A.V. Radyushkin, 
                      Phys. Lett. B {\bf 94}, 245 (1980);
               G. P. Lepage and S. J. Brodsky, 
                      Phys. Lett. B {\bf 87}, 359 (1979).
\bibitem{j-parc} The J-PARC project is found http://j-parc.jp/index-e.html;
                 S. Sawada, Nucl. Phys. A {\bf 782}, 434 (2007);
                 S. Kumano, Nucl. Phys. A {\bf 782}, 442 (2007).
\bibitem{gsi-fair} http://www.gsi.de/fair/index\_e.html.
\bibitem{Zhalov:2000hk}
   D. Zhalov {\it et al.}  (E850 Collaboration),
          AIP Conf. Proc. {\bf 549}, 310 (2002);
   D. Zhalov, 
          Ph. D. Thesis (report UMI-30-36627), 
          Pennsylvania State University (2001).
\bibitem{bf7375} S. J. Brodsky and G. R. Farrar, 
                    Phys. Rev. Lett. {\bf 31}, 1153 (1973);             
                    Phys. Rev. D {\bf 11}, 1309 (1975).        
\bibitem{White94} C. White {\it et al.}, Phys. Rev. D {\bf 49}, 58 (1994).                       
\bibitem{Mueller}
   A. H. Mueller, p.13 in Proceedings of 17th rencontre de Moriond,
            edited by J. Tran Thanh Van 
            (Editions Frontieres, Gif-sur-Yvette, France, 1982) Vol. I.
\bibitem{Brodsky} 
   S. J. Brodsky, p.963 in Proceedings of the 13th International Symposium
            on Multiparticle Dynamics, edited by W. Kittel, W. Metzger, 
            and A. Stergiou (World Scientific, Singapore 1982).
\bibitem{Frankfurt:1994hf}
   L. L. Frankfurt, G. A. Miller, and M. Strikman,
        Ann. Rev. Nucl. Part. Sci. {\bf 44}, 501 (1994).
\bibitem{:2007gqa}
   B. Clasie {\it et al.}, Phys. Rev. Lett. {\bf 99}, 242502 (2007).
\bibitem{Larson:2006ge}
   A. Larson, G. A. Miller, and M.~Strikman,
        Phys. Rev.  C {\bf 74}, 018201 (2006).
\bibitem{chh97} C. E. Carlson, J. R. Hiller, and R. J. Holt,
                    Ann. Rev. Nucl. Part. Sci. {\bf 47}, 395 (1997).
\bibitem{lattice-glue} For example, see
   Y. Chen {\it et al.}, Phys. Rev. D {\bf 73}, 014516 (2006).
\bibitem{lightcone-wave}  
         K.-C. Yang, Nucl. Phys. B {\bf 776}, 187 (2007). 
\bibitem{f0}
       F. E. Close and N. A. T\"ornqvist, J. Phys. {\bf G28}, R249 (2002);
       F. E. Close, N. Isgur, and S. Kumano, Nucl. Phys. {\bf B389}, 513 (1993);
       M. Hirai, S. Kumano, M. Oka, and K. Sudoh, 
                    Phys. Rev. D {\bf 77}, 017504 (2008). 
\bibitem{n-delta-isospin} 
                A. R. Edmonds, {\it Angular Momentum in Quantum Mechanics}
                (Princeton University Press, 1974);
                S. Kumano, Phys. Rev. D {\bf 41}, 195 (1990).
\bibitem{difference} 
       The definition of $n^\mu$ is taken from Ref. \cite{gpv01}
       and it is different from the one in Ref. \cite{diehl03},
       where $n^\mu_- =(1,0,0,-1)/\sqrt{2}$.
       The definition of other variables, $P$, $\Delta$, and $\xi$
       are the same in both articles.
\bibitem{note-n}
       In our actual calculations in Sec .\ref{results}, a different
       one ($n_+^\mu$) is used by considering $(p_a)_z<0$ in the c.m. frame
       of initial nucleons $a$ and $b$.
       The details are explained in Sec. \ref{results}.
\bibitem{js73} 
   H. F. Jones and M. D. Scadron, 
         Ann. of Phys. {\bf 81}, 1 (1973).
\bibitem{bd-book} J. D. Bjorken and S. D. Drell,
                    {\it Relativistic Quantum Fields}
                    (McGraw-Hill, Inc., 1965).
\bibitem{Rarita-Schwinger} 
             S. Gasiorowicz, {\it Elementary Particle Physics}
                      (Jone Wiley \& Sons, Inc., 1966);
             D. Luri\'e, {\it Particles and Fields} 
                      (Jone Wiley \& Sons, Inc., 1968).
\bibitem{h-notation} 
        We denoted the spin-dependent $N \rightarrow \Delta$
        transition GPDs as $\widetilde H_i (x,\xi,t)$ instead of 
        $C_i (x,\xi,t)$ in Refs. \cite{gpv01,fpsv9800} 
        throughout this article in order to use similar notations 
        to the nucleonic ones.
\bibitem{ppg00} M. Penttinen, M. V. Polyakov, and K. Goeke,
                          Phys. Rev. D {\bf 62}, 014024 (2000).
\bibitem{ujs07} M. Ungaro, K. Joo, and P. Stoler,
                in {\it Shapes of Hadrons},
                edited by C. N. Papanicolas and A. M. Bernstein, 
                AIP Conf. Proc. {\bf 904}, 232 (2007).
\bibitem{adler75} S. L. Adler, Ann. Phys. {\bf 50}, 189 (1968);
                            Phys. Rev. D {\bf 12}, 2644 (1975).
\bibitem{parity-electron} N. C. Mukhopadhyay {\it et al.}, 
                            Nucl. Phys. {\bf A633}, 481 (1998);
           K. A. Aniol {\it et al.} (HAPPEX Collaboration),
                        Phys. Rev. C {\bf 69}, 065501 (2004);
           C. Alexandrou, T. Leontiou, J. W. Negele, 
                        Phys. Rev. Lett. {\bf 98}, 052003 (2007).                            
\bibitem{geng-etal-08}   L. S. Geng, J. Martin Camalich, L. Alvarez-Ruso,
                     and M. J. Vicente Vacas, 
                       Phys. Rev. D {\bf 78}, 014011 (2008).
\bibitem{pi-n-delta-calc} 
      L. L. Frankfurt, L. Mankiewicz, and M. I. Strikman, 
                       Z. Phys. A {\bf 334}, 343 (1989); 
      S. Kumano, Phys. Rev. D {\bf 43}, 59 (1991); 3067 (1991);
                 Phys. Rept. {\bf 303}, 183 (1998);
      S. Kumano and J. T. Londergan, Phys. Rev. D {\bf 44}, 717 (1991).
\bibitem{bl} G. P. Lepage and S. J. Brodsky,
                          Phys. Lett. B {\bf 87}, 359 (1979);
             S. J. Brodsky, T. Huang, and G. P. Lepage, 
                          report SLAC-PUB-2540 (1980);
             T. Huang, B.-Q. Ma, and Q.-X. Shen, 
                          Phys. Rev. D {\bf 49}, 1490 (1994).
\bibitem{rho} S. J. Brodsky {\it et al.}, 
                         Phys. Rev. D {\bf 50}, 3134 (1994). 
\bibitem{bnl-e755} G. C. Blazey {\it et al.}, 
                         Phys. Rev. Lett. {\bf 55}, 1820 (1985);
                   B. R. Baller {\it et al.}, 
                         Phys. Rev. Lett. {\bf 60}, 1118 (1988).
\bibitem{pdg08}
      C. Amsler {\it et al.} (Particle Data Group), 
             Phys. Lett. B {\bf 667}, 1 (2008).
\bibitem{feynman} R. P. Feynman, p.91 in
            {\it Photon-Hadron Interactions}
            (W. A. Benjamin, Inc., 1972).
\bibitem{pdg06} Review of Particle Physics 2006 Edition, \\
                    http://ccwww.kek.jp/pdg/2006/pdg\_2006.html.
\end{thebibliography}
\end{document}